


\def\a{\alpha}
\def\b{\beta}
\def\g{\gamma}
\def\d{\del}
\def\e{\epsilon}
\def\m{\mu}
\def\n{\nu}

\def\s{\sigma}
\def\mn{{\mu\nu}}
\def\rarr{\rightarrow}

\def\half{{1\over 2}}
\def\thalf{{3\over 2}}

\def\Order{{\cal O}}
\def\bar#1{\overline{#1}}
\def\bold#1{\setbox0=\hbox{$#1$}%
     \kern-.025em\copy0\kern-\wd0
     \kern.05em\copy0\kern-\wd0
     \kern-.025em\raise.0433em\box0 }
\def\del{\partial}
\def\dslash{\not{\hbox{\kern-2pt $\partial$}}}
\def\Dslash{\not{\hbox{\kern-4pt $D$}}}
\def\Qslash{\not{\hbox{\kern-4pt $Q$}}}
\def\pslash{\not{\hbox{\kern-2.3pt $p$}}}
\def\kslash{\not{\hbox{\kern-2.3pt $k$}}}
\def\qslash{\not{\hbox{\kern-2.3pt $q$}}}
\def\pairof#1{#1^+ #1^-}

\def\ee{\pairof{e}}

\def\L{{\cal L}}
\def\M{{\cal M}}
\def\E{{\cal E}}
\def\D{{\cal D}}
\def\Z{{\cal Z}}
\def\sstw{\sin^2\theta_w}
\def\cstw{\cos^2\theta_w}

\def\msb{{\bar{\ssstyle M \kern -1pt S}}}

\def\long{{\rm long}}
\def\refmark#1{\attach{\scriptstyle [ #1 ] }}

%

%
\pubnum{6453}
\date{April, 1994}
\pubtype{(T/E)}
\titlepage
\title{Spin, Mass, and Symmetry}
\author{Michael E. Peskin\doeack}
\SLAC
\vfil
\centerline{Lectures presented at the XXIst SLAC Summer Institute}
\centerline{Spin Structure in High Energy Processes}
\centerline{Stanford California, July 26--August 6, 1993}
\endpage
\noindent{\bf Table of Contents}
\bigskip
\settabs 8\columns{
\+ 1. Introduction\cr
\+ 2. The Poincar\'e Group \cr
\+   & 2.1. The Rotation Group\cr
\+   & 2.2. The Lorentz Group\cr
\+   & 2.3. Fields under the Poincar\'e Group\cr
\+   & 2.4. Particles under the Poincar\'e Group\cr
\+   & 2.5.  Massless Particles under the Poincar\'e Group\cr
\+ 3. Spin $\half$\cr
\+   & 3.1. Spin $\half$ Lagrangians\cr
\+   & 3.2. Spin Decoupling at Low and High Energy\cr
\+ 4. Spin 1 \cr
\+   & 4.1. Quantum Electrodynamics\cr
\+   & 4.2. Massive Spin 1 Bosons\cr
\+   & 4.3. Examples from the Standard Model \cr
\+ 5. Higher Spin\cr
\+ 6. Spin $\half$ as a Construct\cr
\+   & 6.1. A Model of a Scalar Particle\cr
\+   & 6.2. Addition of Spin     \cr
\+ 7. Spin $\half$ as a Symmetry \cr
\+   & 7.1. The Supersymmetry Algebra\cr
\+   & 7.2. Supersymmetric Dynamics \cr
\+   & 7.3. Spin $\thalf$ and Higher Supersymmetries\cr
\+ 8.  A Fruitful Blend          \cr
\+   & 8.1. The Bosonic String        \cr
\+   & 8.2. Decorated Strings         \cr
\+ 9.  Conclusions               \cr }
\endpage
\chapter{Introduction}

   When the strong interactions were a mystery, spin seemed to be just a
complication on top of an already puzzling set of phenomena.
But now that particle physicists have
understood the strong, weak, and electromagnetic interactions to be
gauge theories, with matter built of quarks and leptons, we recognize
that the
special properties of spin $\half$ and spin 1 particles have taken
central role in our understanding of Nature.  The lectures in this
summer school will be devoted to the use of spin in unravelling detailed
questions about the fundamental interactions.   Thus, why not begin
by posing a deeper question:  Why is there spin?  More precisely,
why do the basic pointlike constituents of Nature carry intrinsic
nonzero quanta of angular momentum?

   \REF\Umez{H. Umezawa, \sl Quantum Field Theory. \rm (North
                     Holland, Amsterdam, 1956).}
     \REF\Weinberg{S. Weinberg, \sl Phys. Rev. \bf 133, \rm B1318,
            \bf 134, \rm B882 (1964).}
   \REF\vanN{P. van Nieuwenhuizen, \sl Phys. Repts. \bf 68,  \rm
                   189 (1981).}
  The nature and realization of spin is one of the deep questions in
quantum field theory.  The subject has great technical complication
and is often relegated to technical treatises or highly specialized
articles.  Some detailed treatments of spin in quantum field theory
are given in refs. \Umez--\vanN.  But, though the technical answers
are often complex, the general ideas of the physics of spin are
of genuine interest to those who would like to understand modern
particle physics.  In these lectures, I would like to give a
broad-brush treatment of this subject, emphasizing its major ideas
and challenging questions.

     Why is there spin?
   Three different kinds of explanatory principles can be brought forth
to answer this question.  These might be called {\it permissive},
{\it a posteriori}, and {\it a priori\/} or {\it constructive\/}
explanations.  Some people are
satisfied with an explanation at any of these levels; others will insist
on the third, strongest type of explanation.  Let us consider each
level in turn.

   A {\it permissive\/} explanation invokes the Totalitarian Principle
 of Physics:  Whatever is allowed, must exist.  Under this philosophy,
we can explain spin by showing that it is a natural consequence of
some general formal structure.  I will review in the next section the
idea, uncovered by Wigner, that the representations of the Poincar\'e
group include naturally include point particles with intrinsic spin.
If such particles are possible, why can't they occur?

  The idea of {\it a posteriori} explanation takes this argument a
step further.  At this level, one still will not claim to understand
why particles have spin, but one argues that, without it, there would
be a disaster.  Such arguments use the Anthropic Principle  that the
world we see must be such that we can live in it.  The Anthropic
Principle gives a particularly strong case for the existence of spin:
Without spin, particles would not obey the Pauli exclusion principle.
But without the stability of Pauli exclusion, matter could avoid
collapse only by finding delicate equilibrium states,  such as that of
the Wigner crystal, which become unstable at high density.
Thus, it would be extremely difficult to build up the ordered
assemblages of matter that are needed to make intelligent life.

  Are these principles satisfying?  Ultimately, this question goes
beyond physics.  It is possible that a Creator envisioned an ordered
Universe and included the ingredients necessary to bring it about.
Linde has argued for another point of view, that the universe contains
as small domains regions in which the laws of physics are realized in
all possible ways.\Ref\Linde{A. D. Linde, \sl Phys. Scripta \bf
               T15, \rm 169 (1987);  A. S. Goncharov,  A. D. Linde,
              and V. F. Mukhanov, \sl Int. J. Mod. Phys. \bf A2, \rm 561
                      (1987).}
Then we inhabit the domain in which we can live.

  However, both of these explanatory principles seem to me much less
compelling that a {\it constructive\/} principle which  explains the
ingredients of Nature as consequences of a grand pattern of symmetry.
The constructive argument for the existence of particles with spin
1 is by now familiar to all particle physicists:  If the equations of
the universe  possess a local gauge symmetry, then to each generator
$Q^a$ of the gauge group, there must correspond a vector field
$A_\mu^a$.  The quantization of this field produces spin 1 particles.
Unfortunately, there is no equally simple and compelling argument
for spin $\half$.

   How close can we come to a constructive argument for spin $\half$?
Can we find a unified explanation for particles of spin $\half$,
spin 1, and perhaps higher spins?
That is the question I will explore in these lectures.  First, of
all, I will build up the basic formalism of spin.  In Section 2, I
will review the general principles which govern particles and fields
with spin; then I will apply these principles successively, in Section
3, to spin $\half$,  in Section 4, to spin 1, and in Section 5, to
spin $\thalf$ and higher.
With this foundation, I will turn in
Sections 6--8 to the question of the origin of spin $\half$,
reviewing three proposals of increasing sophistication.

\chapter{The Poincar\'e Group}

  Any formal discussion of spin must start from the representations of
 the Poin\-car\'e group, the fundamental spacetime symmetry group of
translations and Lorentz transformations.  Any object that lives in
Minkowski space must belong to some representation of the Poincar\'e
group.  By constructing the simplest representations of the Poincar\'e
 group, we will find that intrinsic spin appears in a natural way.

   One subtlety of this discussion will be that particles and fields
transform in different representations of the Poincar\'e group.
In elementary discussions of quantum field theory, one is taught that
there is a direct correspondence between the particle and the field.
For fields with spin, we will see that this correspondence is not
so simple.  In fact, the difficulty in finding the correspondence
between particles and fields for fields of high spin will turn out to
be an essential one which gives a crucial restriction on what fields
can appear in Nature.

  \section{The Rotation Group}

   The generators of the Poincar\'e group are three sets of vectors,
the generators of rotations, boosts, and translations.  We will call
these
$$           J^i \ ,  \quad     K^i \ ,  \quad         P^i \ ,
\eqn\generators$$
respectively.  As a first step toward finding the representations of
this group, we can start with a small, familiar piece, the group
of rotations.

   The generators of rotations obey the commutation relations
$$           \bigl[ J^i, J^j \bigr] = i \e^{ijk} J^k   .
\eqn\rotcomm$$
The representations of these commutation relations are familiar
from any book on nonrelativistic quantum mechanics: They are
the multiplets of spin $j$, the states $\ket{j,j^3}$, with
$j = 0,\half, \ldots$ and $j^3 = -j,\ldots, j$.

   The simplest nontrivial representations are those of with $j = \half$
and $j = 1$. For $j= \half$, we represent
$$       J^i = \half \sigma^i \   ,
\eqn\spinhalf$$
with $\sigma^i$ a $2\times 2$ Pauli sigma matrix.  These generators, and
the $2\times 2 $ rotation matrices built from them, act on 2-component
spinors $\xi_\a$, with $\a = +,-$ corresponding to $j^3 = +\half,-\half$.

   For $j=1$, the representation consists of 3-dimensional vectors $v^i$,
and so the $J^i$ must be represented by $3\times 3$ matrices.  For
example
$$      J^3    =  -i \pmatrix{0 & 1&0 \cr-1 &0 &0 \cr0 &0&0\cr} .
\eqn\spinoneex$$
There is another way to describe this matrix action, as follows:
Consider a system with two 2-component spinors.  The state of this system
is described by a tensor carrying two spinor indices,  $T_{\a\b}$.
  Any such tensor
can be divided into its symmetric and its antisymmetric part.  The
most general $2\times 2$ antisymmetric tensor is proportional to
$ \e_{\a\b}$; this object is invariant to spinor rotations.  The
remaining symmetric  $2\times 2$ tensor has  3 independent components
and transforms, in fact, precisely as the 3-dimensional $j=1$
representation of angular momentum. This decomposition of a $2\times 2$
matrix of spinors into an invariant ($j=0$) and a $j=1$ multiplet is
just the familiar angular momentum decomposition
$$              \half \times \half  = 0 + 1   \ ;
\eqn\halftimeshalf$$
you might recall that the $j=0$ is the antisymmetric combination and
the $j=1$ is the symmetric combination of two spin $\half$ systems.

  This construction generalizes to any $j$.  The multiplet of spin $j$
can always be represented as  a totally
symmetric tensor with $2j$ two-component spinor indices:
$$               \Xi_{\{\a\b\cdots \delta\}} .
\eqn\thetensor$$
It is easy to check that this object has $2j+1$ components, and that
its highest values of $j^3$, given by $\a=\b=\cdots=\delta= +$, is
$j^3=j$.  You can view \thetensor\ as what remains when the lower-spin
components of a general tensor are projected out by contracting
indices with the invariant $\e^{\a\b}$.

   Now that we have  a general picture of the representations of the
rotation group, we can find the representations appropriate to
particles and to fields.  Particles are particular
states of the Hilbert space with localized excitation;
these
 can be classified by
their values of $j$ and $j^3$:
$$           \ket{j,j^3}     \ .
\eqn\particstates$$
Fields are operators which are functions of the space-time position
$x^\mu$.  A general field can be written as a member of a multiplet
of fields
$$                 \Phi_{\{\a\b\g\}}(x) \ ,
\eqn\fieldstates$$
in which, one must remember, a rotation acts both on the spinor
indices and on the spatial position.  Usually, a field with explicit
indices corresponding to spin $j$ will create a particle of
intrinsic spin $j$.  However, this correspondence is not at all
obvious, since the field \fieldstates\ will create eigenstates of
the Hamiltonian with all (half-integer) values of angular momentum.
To understand the correspondence between particles and fields, we must
probe more deeply.

 \section{The Lorentz Group}

  The next step in finding to representations of the Poincar\'e group
is to add the  generators of boosts.  This gives the commutation
relations of the Lorentz group:
$$ \eqalign{ \bigl[ J^i, J^j \bigr] &= i \e^{ijk} J^k  \cr
 \bigl[ J^i, K^j \bigr] &= i \e^{ijk} K^k  \cr
 \bigl[ K^i, K^j \bigr] &= -i \e^{ijk} J^k.  \cr}
\eqn\Lorcomm$$
The appearance of $J^k$ in the last line tells us that the composition
of boosts produces a rotation; this effect is known as the {\it Wigner
rotation}.  Indeed, essentially all of the representation theory
from here on was first formulated by  Wigner.\Ref\WignerL{E. Wigner,
       Ann. Math. \bf 40, \rm 149 (139).}
The minus sign in the last line of \Lorcomm\ tells us that the
rotations generated by $J^i$ and $K^i$ are not four-dimensional rotations
covering a compact space but rather are transformations which span
noncompact spaces---the hyperboloids of Minkowski geometry.

   There is a simple trick for finding the representations of the
commutation relations \Lorcomm.  Let
$$   J_\pm^j = \half \bigl(J^j \pm i K^j \bigr)  \  .
\eqn\Jpmdef$$
Then the generators $J_+^j$ and $J_-^j$ commute with one another and
obey the commutation relations
$$ \bigl[ J_\pm^i, J_\pm^j \bigr] = i \e^{ijk} J_\pm^k
\eqn\Jpmcomms$$
among themselves.  These latter commutation relations are identical to
the
commutation relations of angular momentum.  Thus, we can find
representations of the original Lorentz group relations \Lorcomm\
by choosing a representation for $J_+^j$ of definite
angular momentum $j_+$, choosing a representation for $J_-^j$ of
definite angular momentum $j_-$, and then recombining  these into
$J^j$ and $K^j$ by inverting \Jpmdef:
$$               J^j = J_+^j + J_-^j \ , \qquad
       K^j = -i \bigl(J_+^j + J_-^j\bigr) \ .
\eqn\JKreconstruct$$
We denote this representation as $(j_+,j_-)$; it is a representation
of \Lorcomm\ of dimension $(2j_+ + 1) \times (2j_-+1)$. We can write
the object which transforms in this representation
as a tensor
$$    \phi_{\{\dot \a\dot \b\cdots \dot\g\}
\{\zeta \eta \cdots \theta\}} \ .
\eqn\thephitrans$$
In general, I will place a dot over an index acted on by the generators
$J_+^j$.

   Notice the factor of $i$ in the reconstruction of $K^j$ in
\JKreconstruct.  This means that $K^j$ will not be Hermitian, and so
the representation we have constructed will not be unitary.  This is
the unfortunate but inevitable result of attempting to find a
finite-dimensional unitary representation of a noncompact group action.
Under Hermitian conjugation, $(J_+^j)^\dagger = J_-^j$; thus, the
representation $(j_+,j_-)$ is complex, with
$$     (j_+, j_-) ^* =  (j_-,j_+) \  .
\eqn\jpjmconj$$

\section{Fields under the Poincar\'e Group}

  The remaining generators of the Poincar\'e group, the translation
generators $P^i$, commute with $J^i$ and $K^i$ and with each other,
so it is easy to take them into account.  We can now write general
representations of the Poincar\'e group on multiplets of fields.
To construct these, we set up field with the spinor indices corresponding
 to a representation $(j_+,j_-)$ of the Lorentz group.  We then make
the field a function of $x^\mu$, allowing rotations, boosts, and
translations to have their standard action on this spacetime
coordinate.

  Here are some examples of this construction.  In each case, I would
like to indicate in particular the action of a boost in the $\hat 3$
direction.  To parametrize boosts, I will use the rapidity $y$,
defined by
$$            e^y = \gamma(1+\beta) = \sqrt{{1+\beta\over 1-\beta}}\ .
\eqn\rapiditydef$$
With this notation, a
boost is represented in general as
$$                \exp\bigl[  i \vec y \cdot \vec K \bigr]
\eqn\yisaboost$$
in particular, in successive boosts along the same axis, the rapidities
add.

  The simplest representations of the Lorentz group are those with one
spinor index: $(\half,0)$ and $(0,\half)$.  From the $(\half,0)$
representation, we can built a field $\psi_{\dot\a}(x)$.  Under a
rotation about the $\hat 3$ axis, this field transforms as
$$            \psi \rarr    e^{i\theta \sigma^3/2} \psi \ ,
\eqn\psirottrans$$
that is, as a spin $\half$ object.  Under a boost, it transforms
as
$$   \psi \rarr    e^{y\sigma^3/2} \psi \ ;
\eqn\psiboosttrans$$
this transformation increases the field amplitude if the spin is
parallel to $\hat 3$.  A field in the $(0,\half)$ representation,
$\psi_\a(x)$, has the same transformation under rotations, but the
opposite transformation under boosts.
$$   \psi \rarr    e^{-y\sigma^3/2} \psi \ .
\eqn\psiboosttranstwo$$

  The next example is a field in the $(\half,\half)$ representation,
$V_{\a\dot\a}(x)$.
This field transforms under rotations as $\half\times \half = $
spin 0 $+$ spin 1.  Under boosts in the $\hat 3$ direction, the
various components of $V_{\a\dot \a}$ transform as
$$ (V_{+\dot+},V_{-\dot-},V_{-\dot+},V_{+\dot-}) \rarr
 (V_{+\dot+},V_{-\dot-}, e^y V_{-\dot+},e^{-y} V_{+\dot-}) \ .
\eqn\variousVtrans$$
All of these properties correspond to those  of a field with a 4-vector
index $V^\mu$. Such a field transforms under rotations as a
multiplet $(V^0, \vec V)$---spin 0 plus spin 1---and the combinations
of components
$$       (V^1+iV^2, V^1-iV^2, V^0 + V^3, V^0-V^3)
\eqn\vcomps$$
transform under boosts according to \variousVtrans.

  The last simple representation we consider is that of a field
belonging to the $(1,0)$ representation: $\Phi_{\{\dot\a \dot \b\}}(x)$.
This field is complex, with its complex conjugate belonging to
the $(0,1)$ representation.  The two pieces together give a
structure with 6 real degrees of freedom.  The two
 fields transform under rotations
 as  spin 1.  Under boosts the three components of $\Phi$
 transform as
$$  (\Phi_{\dot+\dot+}, \Phi_{\dot +\dot - },\Phi_{\dot-\dot-})
 \rarr
 (e^y\Phi_{\dot+\dot+}, \Phi_{\dot +\dot - },e^{-y}\Phi_{\dot-\dot-})
\eqn\variousPhitrans$$
All of these properties accord with the identification of $\Phi$ as
the combination  of electromagnetic fields
$$           \E^i = E^i + i B^i   \ .
\eqn\combineEE$$
The field components
$$        (\E^1 + i \E^2,  \E^3,  \E^1 - i \E^2)
\eqn\EEcomps$$
indeed transform as \variousPhitrans.  The conjugate combination
of fields $\bar\E^i =  E^i - i B^i$   belongs to the $(0,1)$
Lorentz representation.

  The last two examples presented familiar vector and tensor fields in a
rather unfamiliar notation.  To connect the formulae given here to
more familiar ones, we should recall Dirac's famous trick for finding
representations of the commutation relations of the Lorentz group.
 Dirac suggested that one find matrices which
satisfy the simpler algebra
$$            \big\{ \g^\m, \g^\n \big\} = 2 g^\mn
\eqn\Diracalg$$
and form the combinations
$$         \Sigma^\mn = {i\over 4} \bigl[ \g^\m, \g^\n\bigr]  \ .
\eqn\sigmadef$$
Then the components
$$    J^i = \half \e^{ijk}\Sigma^{jk} \ , \quad K^i = \Sigma^{oi}
\eqn\diracsboost$$
satisfy \Lorcomm.
In 4-dimensional spacetime, the simplest representations of
\Diracalg\ are $4\times 4$ matrices.
Dirac's trick gives a representation of the Poincar\'e group
as a 4-component field; this is the standard
Dirac spinor  $\Psi$.

A convenient explicit set of $4\times 4$ matrices satisfying
Dirac's relation \Diracalg\ is
$$  \g^\mu  = \pmatrix {0 & \sigma^\mu\cr \bar \sigma^\mu * & 0 \cr} \ ,
\eqn\sigdef$$
with the $2\times 2 $ components
$$  \s^\m = (1, \vec \sigma) \ , \quad
\bar\s^\m = (1, - \vec \sigma)\   .
\eqn\sigdeftwo$$
In this basis, the combinations \Jpmdef\ are given by
$$     J_+^i = \pmatrix{ 0 & \cr & \sigma^i/2 \cr} \ , \qquad
     J_-^i = \pmatrix{ \sigma^i/2 &\cr & 0 \cr} \ .
\eqn\wellfindJpm$$
Thus, the Dirac spinor $\Psi$ is revealed to be a pair of 2-component
fields which transform as a $(0,\half)$ and a $(\half,0)$ under the
Lorentz group:
$$   \Psi = \pmatrix{ \psi_\a \cr \psi_{\dot\a}\cr} \ .
\eqn\decompofPsi$$

  With this notation, the components of $\g^\m$ carry
the indices $\s^\mu_{\a \dot\a}$.  It is quite appropriate to think
of these constant matrices as the Clebsch-Gordon coefficients which
link the $(0,\half)$, $(\half,0)$, and $(\half,\half)$ or 4-vector
representations.  The two sets of $\sigma^\mu$ are not distinct;
they are related by a similarity transformation:
$$   \bar\s^\mu  =  \s^2 (\s^\m)^T \s^2
\eqn\unbars$$
The factor $\s^2$ reflects the complex conjugation relation of
the $(0,\half)$ and $(\half,0)$ representations.
In order to build a field  from $\psi_\a$
which transforms {\it exactly} like a $(\half,0)$,
 one must change the basis for the conjugate of
$\psi_\a$ according to
$$    \tilde \psi_{\dot a} =  ( \psi^\dagger \sigma^2 )_{\dot\a} \ .
\eqn\conjofpsi$$

 Using the invariant $\sigma^\mu$, we can identify the fields
$V_{\a \dot\a }$ and $\Phi{\{\dot\a \dot \b\}}$ about with fields
carrying more familiar combinations of indices.  For the vector
field
$$                 V_{\a \dot\a} =   \sigma^\mu_{\a\dot \a}  V_\mu  \ .
\eqn\vident$$
The electromagnetic field strength is usually written as an antisymmetric
tensor  $F_\mn = (\del_\m A_\n - \del_\n A_\m)$.  Then we can write
$$    \Phi_{\dot\a\dot\b} = \s^\m_{\a\dot\a} \s^\n_{\b\dot\b} \e^{\a\b}
                  F_\mn \ .
\eqn\phiident$$
Notice that the tensor $\Phi$ is indeed required to be symmetric
as a consequence of the antisymmetry of the other elements in
\phiident.

\section{Particles under the Poincar\'e Group}

To describe the transformations of particles under the Poincar\'e
group, we use a somewhat different language.  While fields are
operators which carry indices, particles are states in the Hilbert
space of the quantum field theory.  The transformation of a field
need not be unitary, but transformations of states in  the
Hilbert space must be.  It is thus useful to represent the various boosts
and rotations of a given particle by the actions of these abstract
unitary transformations.

If the particle has mass, it is most convenient to begin from its rest
frame.  In this frame, a particle of spin $s$ forms a multiplet of
$(2s+1)$ states
$$            \ket{\vec p = 0; s \, s^3}
\eqn\restket$$
which transform into one another under rotations.
The boosts of these states can be defined as
$$  \ket{\vec p; s\, s^3} = \Lambda(\vec p) \ket{\vec 0; s\, s^3}\ ,
\eqn\boostket$$
where $\Lambda(\vec p)$ is the unitary transformation which implements
the boost.

Since boosts and rotations do not commute, we profit from being very
careful in defining the order of the boosts and rotations that lead to
a given states.  For relativistic particles, it is often most
convenient to quantize the spin along the direction of motion.  In this
system, states are labeled by their {\it helicity}, their
  spin projection along the direction
of motion.  If $\hat p$ is a unit vector parallel to $\vec p$, the
helicity is
$$     \lambda = \vec s \cdot \hat p   \  .
\eqn\lambdadef$$
The wonderful properties of this representation are explained in a
classic paper of Jacob and Wick.\Ref\JW{M. Jacob and G. C. Wick, \sl
Ann. Phys. \bf 7, \rm 404 (1959).}  To write a
state explicitly in the helicity representation, start from a specific
spin state in the rest frame,
 boost from a rest parallel
to the $\hat 3$ axis, and then rotate to bring the momentum $\vec p$
into its correct orientation.  If the orientation of $\vec p$ is
given by polar and azimuthal angles $\theta$ and $\phi$, the state of
helicity $\lambda$ is defined from the rest frame state by
$$  \ket{p,\theta,\phi;\lambda} = e^{-i\phi J^3}e^{-i\theta J^2}
e^{i\phi J^3} \Lambda(p\hat 3) \ket{\vec 0; s\, s^3 = \lambda}\ .
\eqn\lambdaform$$
Notice that the helicity $\lambda$ appears only in the rest frame
state.  Helicity is invariant under spatial rotations and under boosts
parallel to the direction of motion.

The multiplet of states of the form \lambdaform\ form a unitary
representation of Poincar\'e group.  This representation is
infinite-dimensional.  As we have noted, that is a necessary property
if we insist that the group action is unitary.  But this means that
there is no automatic relations between
the transformation properties of
particles and fields.

The simplest way to make a correspondence between the  particle and field
transformations is to connect the field with the particle state that it
creates or destroys.
  For low spin, this is straightforward.  The free scalar field
$\phi(x)$ creates and destroys
scalar particles.  The Fourier transform
$\tilde{\phi}(p)$ precisely  destroys particles with
momentum $p$.  For spin-$\half$, there is a similar relation:  the free
Dirac field $(\psi_\alpha(x), \psi_{\dot\alpha}(x))$ destroys
spin-$\half$ particles and creates their antiparticles according to
relations
$$          \bra{0} \psi_\alpha(x) \ket{p,\lambda} = u^\lambda_\alpha(p)
e^{-ip\cdot x} \ ,
\eqn\destroyDirac$$
where the right-hand side is a solution to the free Dirac equation.
Note that in this case half of the components correspond to particles
destroyed, while the other half correspond to antiparticles created.

However, beginning with spin 1, problems arise in this identification.
A free vector field $V^\mu(x)$ creates a particle polarized in the
direction $\mu$.  This is confusing if $\mu = 0$, since a vector
particle has only three polarization states, corresponding in the
rest frame to the three spatial directions. If we had a fourth
polarization state of a vector particle, its inner product with the
other states would need to conform to the requirements of relativistic
invariance, and we would find
$$   \VEV{p,\mu \bigm| p', \nu}  = - g^{\mu\nu}\delta(p-p')\ .
\eqn\negamm$$
This is a negative inner product---negative probability---for $\mu=\nu
= 0$.  This mismatch persists for the spin-$\thalf$ field
$\psi_{\mu\alpha}$, and gets worse for fields with multiple 4-vector
indices.

\section{Massless Particles under the Poincar\'e Group}

The mismatch between particle and field degrees of freedom, which is
already a problem for massive particles, becomes even worse for
massless particles.  To understand the new complication, we should
think a bit more about the invariances of the helicity.

\FIG\ChangeHelic{The process which changes the helicity of a
massive particle.}
\FIG\MinRep{The minimal representation of the Poincar\'e
group in the case of a massless particle.}

For a massive particle, helicity is not invariant to all operations of
the Poincar\'e group.  It is easy to see that the massive particle can
be boosted to rest, and then boosted into any other direction, allowing
an arbitrary change in its helicity (Fig. \ChangeHelic). However,
this pathway is not available for a massless particle, which can never
be boosted to rest.  In fact, for a massless particle, the helicity is
a Poincar\'e invariant.  This means that massless particles live in
extremely small representations of the Poincar\'e group.  A typical one
is shown in Fig. \MinRep.  It consists of a particle in a state
of definite helicity $\lambda$, which may be boosted to an arbitrary
lightlike momentum, and its conjugate under $CPT$, which is an
antiparticle of helicity $-\lambda$.

Massless particles are created and destroyed by fields which obey
massless wave equations.  Thus, one might ask, which component of the
field creates the particle?  To answer this question rigorously, one
must perform a careful analysis of the field equation.  Here, I will
give a partial answer to the question using a shortcut which involves
dimensional analysis.  To begin, recall that the matrix element through
which a field destroys a particle is dimensionless in the case of an
integer-spin field and proportional to $|p|^{1/2}$ in the case of a
half-integer spin field:
$$   \bra{0}\Phi(x) \ket{p,\lambda}  = \bigg\{ \matrix{ \epsilon\sim 1\cr
          u(p) \sim |p|^{1/2} \cr} \bigg\} e^{-ip\cdot x}\ .
  \eqn\psiformdim$$
Assume that $p$ is parallel to the $\hat 3$ axis.
We now write the state $\ket{p,\lambda}$ as the boost $\Lambda
= e^{iyK^3}$ of a state at
lower momentum $p'$.  Since the vacuum is boost invariant, we can
 rearrange the matrix element as follows:
$$ \eqalign{
  \bra{0}\Phi(x) \ket{p,\lambda} & = \bra{0} \Phi(x) \Lambda
\ket{p',\lambda}\cr
  &= \bra{0}\Lambda^{-1} \Phi(x) \Lambda\ket{p',\lambda}\ .\cr}
 \eqn\changeLam$$
 The dependence of the matrix element on $p$ is now contained in the
 transformation law of the operator, and we can work this out using the
 formulae of Section 2.3.

 Consider first a spinor field with an undotted index.  From eq.
 \psiboosttrans, we can read the transformation law
$$  \Lambda^{-1} \psi_\a \Lambda =  \bigl(
e^{-y\sigma^3/2}\bigr)_{\a\b} \psi_\b .
\eqn\psitratwo$$
This expression is proportional to $|p|^{-1/2}$ for $\a = +$, and to
$|p|^{1/2}$ for $\a = -$.  Only the second relation agrees with
dimensional analysis.
If the particle
were massive, the amplitude for the $\a = +$
component to destroy a fermion could consistently have the
 form $|m^2/p|^{1/2}$ at
large $p$; however, for a massless particle, this form is not
available.
We conclude that $\psi_\a$ destroys only
{\it left-handed} massless spin-$\half$ particles.  By a similar
argument, we would find that this field can also create their
right-handed antiparticles.  Since $\psi_{\dot\a}$ has just the
opposite transformation property under boost, we would find that a
fermion field with a dotted index is associated with {\it right-handed}
massless fermions and left-handed antifermions.

  A similar argument can be made for the matrix element for a vector
field $A_{\a\dot \a}$ to destroy a vector particle.  Applying
\changeLam\ and the transformation law given in  \variousVtrans,
we find
$$
  \bra{0} A_{\a\dot \a}(x) \ket{p,\lambda}  =
      \cases{ e^y \sim p   &  $\a\dot\a = -\dot +$\cr
              1   &  $\a\dot\a = +\dot +, -\dot -$\cr
       e^{-y} \sim p^{-1}   &  $\a\dot\a = +\dot -$\cr} \ .
\eqn\Amatrixrel$$
Only the middle relation is consistent with dimensional analysis.
Thus, $A_{\a\dot\a}$ destroys, and creates, states with helicity
$\lambda = 1$ and  $-1$, but not $\lambda = 0$.

 For higher-spin fields,
this dimensional analysis argument allows more possibilities, and
one must work out the   explicit consequences of the equations of
motion to exclude some of these.  The general conclusion is that
only the field components which create maximal helicity have one-particle
matrix elements.  For the spin-2  field, for example, the field
components which create massless particles are
$$        g_{+ + \dot + \dot +}  \ , \quad  g_{- - \dot - \dot -}\ .
\eqn\gcomps$$
These create and destroy particles of helicity $\pm 2$.

  Up to this point, we have only addressed the question of which
matrix elements can and cannot be zero, on general principles.  It is a
separate question to write a set of equations of motion which lead to
the correct one-particle matrix elements of fields, and which give
these fields a consistent set of interactions.  To study that question
we will consider a series of specific examples, beginning with spin
$\half$ and working upward.

\chapter{Spin $\half$}

  Systems with spin $\half$ provide the simplest examples in which
there is a nontrivial relationship between the quantum fields and the
particles they create and destroy.  Many of the complications we will
find with spin 1 and higher are absent here, but nevertheless, the
equations of motion of spin $\half$ fields have many interesting
features which are dictated by Lorentz invariance.  In addition, the
most important  particles of the standard model---the quarks and
leptons---have spin $\half$, and many of the fundamental questions we
have about these particles are posed most clearly in a language
which appreciates the constraints given by space-time symmetries.

\section{Spin $\half$ Lagrangians}

The easiest way to write a set of Lorentz-covariant field equations is
to derive these equations
 from a Lorentz-invariant Lagrangian.  It is easy to
construct such Lagrangians:  If we begin with fields which carry
dotted and undotted spinor indices, Lorentz-invariance is guaranteed
if we contract all indices of each type.

 As an example, we can construct the Lagrangian for a spin $\half$
field $\psi_\a$.  This Lagrangian should involve the field $\psi_\a$,
is Hermitian conjugate $\psi^\dagger{}_{\dot \a}$, and at least one
spatial derivative $\d_\mu$.  By using the invariant $\bar\sigma^{\mu
\dot\a \a}$ to convert the vector index to spinors, we can contract
all the indices by writing
$$  \L =   \psi^\dagger{}_{\dot\a}\, i \bar\sigma^{\mu\dot \a \a} \d_\mu
              \psi_\a  \ .
\eqn\firstfermL$$
This is the simplest possible spin $\half$ Lagrangian, involving a
2-component, not a 4-component, field.  In a moment, I will show
how to reconstruct the familar Dirac Lagrangian from this starting
point.

  The field equation following from the Lagrangian \firstfermL\ is
$$       i  \bar\sigma^{\mu\dot \a \a} \d_\mu
              \psi_\a  = 0 \ .
\eqn\firstWeq$$
This is the {\it Weyl equation}.  Multiplying on the left by
$i\sigma^\nu_{\a\dot\a}\d_\nu$ and using $\sigma^\nu \bar\sigma^\mu
= g^{\mu\nu}$, this equation becomes
$$           \d^2 \psi_\a = 0 \ .
\eqn\secondW$$
Thus, the Weyl equation is an equation for massless particles.
However, not every massless wave function satisfies \firstWeq.  If we
look for solutions to \firstWeq\ of the form of a plane wave,
$$   \psi = u(p) e^{-ip\cdot x} \ ,
\eqn\solutpsi$$
where $u(p)$ is a 2-component constant vector, this vector satisfies
$$     \bar\sigma\cdot p\, u(p) = (p^0 + \vec \sigma \cdot \vec p) u(p)
                = 0 \ .
\eqn\barsigsol$$
This equation implies that $u(p)$ is proportional to a 2-component
spinor which is left-handed with respect to the direction of motion.
If $\xi_\a$ is a spinor normalized to $||\xi|| = 1$, then
$$     \psi_\a = \sqrt{2E}\, \xi_\a  e^{-ip\cdot x} \ .
\eqn\psisoltwo$$

  Along with the Lagrangian \firstfermL, there is another equally
simple Lagrangian involving the spin $\half$ field with a dotted
index, $\tilde \psi_{\dot\a}$:
$$  \L = \tilde \psi^\dagger{}_{\a}\, i\sigma^{\mu \a\dot \a} \d_\mu
          \tilde\psi_{\dot\a}  \ .
\eqn\secondfermL$$
This Langrangian implies
the field equations equation similar to \firstWeq\ with
$\bar\sigma^\mu$ replaced by $\sigma^\mu$, leading to solutions which
are right-handed with respect to the direction of motion.  However,
this Lagrangian is not an alternative to \firstfermL; instead, it is
identical.  We may replace $\tilde\psi$ with $\psi^\dagger$ according
to \conjofpsi:
$$    \tilde \psi =  ( \psi^\dagger \sigma^2 )^T \ ; \quad
           \tilde\psi^\dagger=  (\sigma^2 \psi)^T \ .
\eqn\conjofpsi$$
Integrate by parts, and cancel the minus sign from this manipulation
against the one obtained by interchanging the order of the fermion
fields.  Finally, use the identity \unbars.  We find that
\secondfermL\ is
transformed into precisely into \firstfermL.

   Both of these forms of the Weyl Lagrangian may be compared with the
standard  Dirac Lagrangian which describes electrons in quantum
electrodynamics:
$$  \L = \bar \Psi i\gamma^\mu (\d_\mu + i e A_\mu) \Psi - m \bar
  \Psi \Psi                    \ .
\eqn\DiracL$$
Replace the four-component Dirac spinor by two two-component spinors
with undotted indices:
$$     \Psi = \pmatrix{\psi_\a\cr \psi_{\dot \a}\cr}
          = \pmatrix{\psi_{1\a} \cr (\psi^\dagger_{2}\s^2)_{\dot \a}\cr}
                    \  .
\eqn\Psidecomp$$
The Dirac Lagrangian is rewritten as follows:
$$\eqalign{
  \L =   \psi_1^\dagger{}_{\dot\a}& i \bar\sigma^{\mu\dot \a \a}
 ( \d_\mu + ie A_\mu)
              \psi_{1\a} +
\psi_2^\dagger{}_{\dot\a} i \bar\sigma^{\mu\dot \a \a}
 (\d_\mu -ieA_\mu)
              \psi_{2\a} \cr  &
  -i m \epsilon^{\a\b} \psi_{1\a} \psi_{2\b} + i m \e^{\dot\a\dot\b}
                        \psi_{1\a}^\dagger \psi^\dagger_{2\b} \ .
  \cr}
\eqn\nextDirac$$
If we ignore the terms proportional to the electron mass, the
Lagrangian splits into two pieces, one for the left-handed
electron and its right-handed antiparticle, and one for the left-handed
positron and its antiparticle, the right-handed electron.  Notice
that the sign of the charge has changed in the second term precisely
in accord with this interpretation.  The mass term is revealed in the
second line of \nextDirac\ to be a  Lorentz-invariant mixing
of the left-handed and right-handed components.

  The structure of eq. \nextDirac\ is very simple; thus, it is
straightforward to generalize it.  In fact, we can immediately write
down the most general Lagrangian for fermions interacting with
vector bosons.  For reasons I will discuss in the next section,
vector bosons are necessarily gauge bosons and are associated with the
generators of a symmetry group.  If we accept this for the moment,
it makes sense to represent the most general collection of fermions
as a collection of two-component fields  $\psi_{a\a}$
on which the gauge symmetries act.  Write the infinitesimal form of this
transformation abstractly as:
$$      \psi_{a\a} \rarr       (1 + i \theta^A T^A)_{ab} \psi_{b\a} \ .
\eqn\psitransgen$$
The gauge symmetry implies that gauge fields couple to fermions
through the {\it covariant derivative}
$$   D_\mu \psi_{a\a} = (\d_\mu \delta_{ab} - i g_A A_\mu^A T^A_{ab})
              \psi_{b\a} \ .
\eqn\covdiv$$
Then the most general Lagrangian for massless fermions has the form
$$\L=  \psi^\dagger_a i \sigma \cdot  D
              \psi_{a}    \ .
\eqn\mostgenLf$$
A mass term for these fermions has the general form
$$\Delta\L = - i M^{ab} \e^{\a\b}  \psi_{a\a} \psi_{b\beta} + h.c.
 \eqn\genmass$$
Notice that the product of two fermion fields is doubly antisymmetric,
since it pick up a minus sign from interchanging the fermion
operators and another from interchanging the indices $\a$ and $\b$.
Thus, the mass matrix $M^{ab}$ is symmetric.

The presentation \mostgenLf, \genmass\ of the fermion Lagrangian
brings us immediately to the most fundamental questions about
elementary fermions.  To write the kinetic energy term \mostgenLf,
we need only the most basic information about these fermions:
how many are there, and how are they organized into representations
of the gauge symmetry group?  To write \genmass, we need to know
how these fermions link up to acquire mass.  Note that these linking
terms often imply breaking of the underlying gauge invariance.
For example, in the electron mass term in \nextDirac, the left-handed
electron field $\psi_1$
is a member of weak isospin doublet, while the left-handed
positron field $\psi_2$ is an isospin singlet.  This brings us directly
to the mystery of what agent breaks this symmetry in order to allow
the mixing of these components.

\section{Spin Decoupling at Low and High Energy}

The Weyl or Dirac Lagrangian dictates a certain relation between the
spin of the fermion and its orbital motion.  To understand this relation,
it is useful to work through some  examples of fermion motion
and its influence on the fermion spin.  The limit of high energy
is especially simple.  In this limit,
the mass terms in the Lagrangian become
irrelevant, and  the Lagrangian decouples into terms involving fermion
components of definite helicity.  Notice that the coupling to
vector fields separates in exactly the same way and also
conserves helicity   in this limit.

Another especially simple limit is that of low energy, in which the
fermion's momentum is small compared to its mass.
In this limit, the effects of relativity become unimportant
and a fermion looks to a good approximation like a scalar particle.
The spin decouples up to effects of order $1/m$.  To see this
explicitly, we manipulate the Dirac equation as follows: Begin
from the equation of motion of the Dirac Lagrangian  \DiracL, in the
form
$$        ( i \gamma\cdot D - m)\Psi = 0 \ , \qquad {\rm where}\qquad
                 D_\mu = \d_\mu - i g A_\mu \ .
\eqn\Diraceqqq$$
Multiply on the left by $(-i\gamma\cdot D - m)$; this gives
$$ \Bigl(  \half\big\{\gamma^\m,\gamma^\n\big\}  D_\m D_\n +
 \half\bigl[\gamma^\m,\gamma^\n\bigr]  D_\m D_\n +  m^2 \Bigr)
           \Psi = 0 \ .
\eqn\Diractwo$$
Now apply \Diracalg\ and  \sigmadef, and simplify the second term
using the antisymmetric relation
$$    \bigl[D_\m,D_\n\bigr]  = -ig F_{\mn}  \ .
\eqn\fidentity$$
This  converts \Diractwo\ into
$$  \Bigl(  D^2 -  g  \Sigma^\mn F_\mn  + m^2 \Bigr) \Psi = 0 \ .
\eqn\finalDirac$$
This last equation is similar to the Klein-Gordon equation, and it is
easy to infer from it the Schr\"odinger equation which  gives its
nonrelativistic limit.  The nonrelativistic Hamiltonian is
$$    H = m - {1\over 2m} (\vec D)^2 + g A^0 - {g\over m} \vec \sigma
            \cdot \vec B   \ .
\eqn\theDHam$$
The first term which involves the spin is also suppressed by an
explicit factor of $1/m$.

  \FIG\EEMM{Kinematics of the process $\ee\rarr \pairof{\mu}$.}

  To describe how the Dirac equation interpolates between these limits,
we will consider a specific practical example, the quantum
electrodynamics
cross section for
the reaction $\ee \rarr \pairof{\mu}$.  The kinematics of the process
are shown in Fig. \EEMM.  To be specific, we will consider
 the annihilation of a left-handed electron
with a right-handed positron, assigning the collision axis in the
direction of the electron motion to be the $\hat 3$ axis.
The electron and positron produce a virtual photon with
spin 1 and $J^3 = -1$ which eventually reforms into a muon pair.
The  differential cross section for this process is easily worked
out from Feynman diagrams.  I will write the result of this calculation
in a suggestive notation.

  In the low energy limit, the muons are produced in an $S$-wave.
Thus, their momenta  are distributed isotropically.  The angular
momentum of the virtual photon must be carried by the muon spins, and
these are approximately decoupled from the   orbital motion.
Then the final muons both have spin $S^3 = -\half$.   In the basis
of $s^3$, we can write the scattering amplitude as
$$ \M(e^-_L e^+_R \rarr \m^-\m^+) =
   -2e^2  \pmatrix{0 & 0 \cr 0 & 1\cr} \ ,
\eqn\theeemat$$
 where the rows of the matrix denote the spin components $S^3 = +\half$
and $S^3 = -\half$ for the $\mu^-$ and the columns denote the spin
components of the $\mu^+$.

  To discuss the transition to high energy, it is convenient to rewrite
this scattering amplitude in a basis of helicity states.  These are
related to the states of definite $S^3$ by a rotation:
$$\eqalign{
 \ket{\mu^-,S^3 = -\half}  & =
 \cos{ \theta\over2} \ket{\mu^-,\lambda = -\half} - \sin{\theta\over 2}
 \ket{\mu^-,\lambda = +\half} \cr
 \ket{\mu^+,S^3 = -\half}  & =
- \cos{\theta\over2} \ket{\mu^+,\lambda = +\half} - \sin{\theta\over2}
 \ket{\mu^-,\lambda = -\half}\ . \cr}
\eqn\helicitybase$$
In the basis of helicity states, the matrix \theeemat\ becomes
$$ \M(e^-_L e^+_R \rarr \m^-\m^+) =
   e^2  \pmatrix{-\sin\theta & (1-\cos\theta) \cr
   (1+\cos \theta)& \sin\theta\cr} \ .
\eqn\theffmat$$
The matrix elements are proportional to the elements of the spin-1
rotation matrices $d^1_{\lambda\lambda'}(\theta)$, as required by the
general results of Jacob and Wick.\refmark\JW

  The expression \theffmat \ can be directly compared to the
high energy limit of the scattering amplitude for $\ee\rarr
 \pairof{\mu}$.  In that limit, we find
$$ \M(e^-_L e^+_R \rarr \m^-\m^+) =
   e^2  \pmatrix{ \Order(m_\mu/E) & (1-\cos\theta) \cr
   (1+\cos \theta)& \Order(m_\mu/E)\cr} \ .
\eqn\theggmat$$
The elements which conserve helicity have the same form as in \theffmat,
while the elements which violate helicity conservation go to zero
as the muons become relativistic.  This matrix element leads to the
unpolarized cross section
$${  d\sigma\over d\cos\theta} = {\pi
\alpha\over 2s}\bigl[ (1+\cos\theta)^2
          + (1 - \cos\theta)^2 \bigr]
\eqn\theQEDform$$
which is familiar from the phenomenology of $\ee$ annihilation
at energies well below the $Z^0$.  More generally, the appearance of
$(1 + \cos\theta)^2$ and $(1-\cos\theta)^2$ angular distributions
in $\ee$ annihilation display the constraint of helicity conservation
at high energy and correlate angular distributions to the couplings
of the various helicity states.

\chapter{Spin 1}

After this taste of the dynamics of spin $\half$ fields, we move on
to a discussion of spin 1.   Spin 1 is the first case in which the
mismatch between field components and physical particles becomes
a serious problem.  In this section, I would like to explain how
this problem is resolved for massless and for massive fields.
The explanation has a surprising number of subtleties, but it also
reveals some interesting physical consequences.

\section{Quantum Electrodynamics}

The most familiar spin 1 particle is the photon.  Since the photon is
massless, one might think that it would be especially difficult to
treat in quantum field theory.  And yet, all of the problems of
principle of building a quantum theory of photons are automatically
answered
in quantum electrodynamics.  Let us review how this happens.

  The Lagrangian of free photons is given by the expressions
$$ \L = - {1\over 4} (F_\mn)^2 = + {1\over 2} A_\mu \bigl(
 \d^2 g^\mn - \d^\m \d^\n\bigr) A_\nu  \ ,
\eqn\theLagforphotons$$
which leads directly to Maxwell's equations.  It is a standard
result of undergraduate physics that the propagating solutions of
Maxwell's equations satisfy
$$    \vec\nabla \cdot \vec E = \vec \nabla \cdot \vec B = 0 \ .
\eqn\transvph$$
A typical solution of Maxwell's equations, propagating in the $\hat 3$
direction, is given by taking the
real and imaginary parts of the relation
$$   \vec E + i \vec B  = (\hat 1\pm i \hat 2) e^{-ip\cdot x}
\eqn\vecforE$$
with $\vec p \parallel  \hat 3$, $p^0 = |\vec p|$.  These are plane
waves, propagating at the speed of light, with helicity $\lambda =
\pm 1$.  In agreement with the dimensional analysis argument
in \Amatrixrel, there is no propagating plane wave solution to
Maxwell's equation with helicity $\lambda = 0$.

  In quantum electrodynamics, we represent the photon field by a
propagator
$$          \VEV {A^\mu(k) A^\nu(-k)} =   {-i g^\mn\over k^2}
\eqn\photprop$$
which apparently contains all field components.  This looks
paradoxical, for two reasons.  First, as we have just discussed,
the helicity zero components of the photon are not associated
with propagating waves.  Second, the expectation value
in \photprop \ seems to indicate an incorrect quantum mechanics.
{}From \photprop, one can straightforwardly derive the identity
$$ \sum_\e
  \bra{0} A^\mu(x) \ket{k,\e} \bra{k,\e} A^\nu \ket{0}
         = - g^{\mn}
\eqn\wrongmetric$$
for the matrix element  of  the vector field between one-particle
states and the vacuum.  If the norms of states in Hilbert space are
positive, this quantity should be positive, but the $\m=\n=0$ element
of \wrongmetric \ is negative.  We encountered this pathology earlier,
in eq. \negamm, and avoided it there only by forbidding bosons
with timelike polarization.  However, when we work with
\photprop, we must necessarily include both bosons with helicity zero
and those with timelike polarization in our
formalism.

    Fortunately, quantum electrodynamics magically resolves both of
these problems.  The crucial element required is the fact
that the photon field couples to a conserved current, the current $j^\m$
of electric charge.
 It is important to note that Maxwell's equations would
be inconsistent if the charge current were not conserved:  In
relativistic form, Maxwell's equations read
$$    \d_\mu F^\mn = ej^\nu \ .
\eqn\conservMax$$
Thus, simply by applying $\d_\nu$ to this equation and using the
fact that $F^\mn$ is antisymmetric, we find
$$            \d_\nu  j^\nu = 0 \ .
\eqn\conservj$$
Alternatively, one can argue that a local gauge symmetry can only be
built in a theory with a perfect global symmetry.

  \FIG\OnePhot{Single photon emission in quantum electrodynamics.}

   The conservation of the current $j^\mu$ constrains the states that
can be produced in quantum electrodynamics processes.  To see this,
consider the matrix element for single photon emission, shown in Fig.
\OnePhot, and analyze this matrix element for a photon emitted parallel
to the $\hat 3$ axis.   If we pull the photon out of the vertex function,
as shown in the figure, we see that it couples to the current $j^\mu$:
$$  i \M = i\M^\mu(q) \e^*_\mu(q) = -ie\VEV{j^\mu(q)}  \e^*_\mu(q)\ .
\eqn\mpull$$
Current conservation imposes the condition
$$  q_\mu \VEV{j^\mu(q)} = 0 \ .
\eqn\currentconsq$$
In this situation, $q^\mu = (q,0,0,q)$, so  this relation implies
$\M^0 = \M^3$. If we take account of the negative norm \wrongmetric\
of time-like polarized photon states, we find a probability of
photon emission proportional to the Lorentz-invariant combination
$$       |\M^1|^2 + |\M^2|^2 + |\M^3|^2 - |\M^0|^2 \ .
\eqn\sumofM$$
Only the first two terms of this expression correspond to physical
propagating photons.  But we now see that the other two terms of this
expression are irrelevant, since they cancel precisely due to the
constraint of gauge invariance.

It is amazing that the various unphysical
aspects of the formalism work together with one another to make this
cancellation occur.  The organizing principle is gauge invariance.
In this discussion, I have made a particular choice of gauge, the
Feynman gauge.  With other choices, for example, the Coulomb
gauge,\Ref\BjD{J. D. Bjorken and S. D. Drell, \sl Relativistic Quantum
Fields, \rm Chapt. 14.  (McGraw-Hill, New York, 1965).} one can
work directly with a Hilbert space which contains only the physical
photon degrees of freedom, at the cost of manifest Lorentz invariance.

 This same cancellation mechanism
also holds in non-Abelian gauge theories with massless gauge bosons.
Again, it is organized by the requirement that the current associated
with the gauge symmetry be (covariantly) conserved.
In the non-Abelian case, there are additional pairs of unphysical
positive and
negative norm states---the Fad'deev-Popov ghosts---which enter the
cancellation.  The detailed proof of this cancellation
is rather technical; it has been set out clearly (for theorists) by
Taylor\Ref\Taylorbk{J. C. Taylor, \sl Gauge Theories of Weak
                  Interactions. \rm  (Cambridge University Press, 1976).}
 and by Kugo and Ogima.\Ref\KandO{T. Kugo and I. Ojima, \sl Supp.
          Prog. Theor. Phys. \bf 66, \rm 1  (1979).}

\section{Massive Spin 1 Bosons}

  The questions that we discussed in the previous section become more
intricate when we consider massive spin 1 bosons.  For massive particles,
all (spacelike) helicity states should be physical and correspond to
propagating modes.  However, in a covariant formalism, the timelike
component of the vector field $A^\mu$ still must create states of
negative norm.  How can these conflicting demands be satisfied?
For simplicity, I will consider only the case of a single massive
vector boson, without the complications of non-Abelian couplings.

  For this case of a single massive spin 1 boson, there are two
solutions known in the literature.  The first is given by adding a mass
term to the  Lagrangian of quantum electrodynamics, to produce the
St\"uckelberg Lagrangian,
$$  \L = - {1\over 4}(F_\mn)^2  + \half m^2 A^\mu A_\mu \ .
\eqn\StuckL$$
This Lagrangian has a very simple classical theory.  The field equation
is
$$  \d_\mu F^\mn + m^2 A^\nu = 0 \ .
\eqn\StuckEq$$
Applying $\d_\nu$ to this equation, we find
$$    m^2 \d_\nu A^\nu = 0 \ ;
\eqn\dnuofA$$
then the timelike component of $A^\mu$ vanishes.
The remaining components of $A^\mu$ satisfy the massive field equation
$$         (\d^2 + m^2) A^\nu = 0 \ .
\eqn\massiveAS$$
The propagating solutions to this equation have the form
$$          A^\mu  = \e^\mu(p) e^{-ip\cdot x} \ ,
\eqn\Asolvecl$$
with $p^2 = m^2$.  Eq. \dnuofA \ imposes the constraint $p\cdot \e(p) =
0$.  The solutions satisfying this constraint correspond to three
spacelike polarizations.  In the quantum theory, by an argument
similar to that of \mpull, the timelike polarization is not produced
from a conserved current.  Unfortunately, the most familiar massive
spin 1 bosons in Nature, the $W$ and $Z$ bosons, couple to currents
such as the weak isospin current
which correspond to broken symmetries.
In this case, the St\"uckelberg strategy breaks down.

  The alternative to this strategy is to construct massive spin 1 bosons
from massless gauge bosons by spontaneously breaking the gauge symmetry.
This strategy, which was discovered by Higgs, Kibble, Guralnik, Hagen,
Brout, and Englert, is now generally known as the {\it Higgs mechanism}.
In its simplest formulation, one would add to the Lagrangian of
electrodynamics an electrically charged scalar field $\varphi$.
$$  \L = - {1\over 4}(F_\mn)^2  + D_\mu \varphi^\dagger D_\mu \varphi
      - V(|\varphi|^2) \ ,
\eqn\HiggsL$$
where the covariant derivative $D_\mu$ is given by $D_\mu = (\d_\mu -ig
A_\mu)$, as in eq. \covdiv.
 The function
$V$ is a potential energy for the field $\varphi$.  If it becomes
energetically favorable for $\varphi$ to obtain a vacuum expectation
value
$$            \VEV{\varphi} = {1\over \sqrt{2} }\, v \ ,
\eqn\varphivacev$$
then the second term in \HiggsL  leads to
$$ D_\mu \varphi^\dagger D_\mu \varphi \rarr \half g^2 v^2 A^\mu A_\mu
             \ ,
\eqn\theHmass$$
which is a mass term with
$m = gv$.  In this way, we recover the St\"uckelberg mass
term, but in a theory with an underlying symmetry structure.

  This structure will become crucial when we try to answer more
detailed questions about the nature of this massive spin 1 field.
Here are two:  Set up the kinematics of boson emission as in
the discussion of eq. \mpull, with the boson moving parallel to the
$\hat 3$ axis.
 In this massive case, $A^3$  creates
physical states, so it is no longer obvious that the negative metric
states created by  $A^0$ will be exactly cancelled.  How is this
guaranteed?  In addition, the new physical states created by $A^3$
have their own difficulties.  If the massive boson is emitted at
rest, the new states have polarization vector $\e^\mu(q) = (0,0,0,1)$.
The boost of this vector to momentum $q$ is
$$   \e^\mu_\long(q) = \bigl( {q\over m},0,0,{E\over m}\bigr) \ ,
\eqn\evector$$
where $E^2 = q^2 + m^2$.  In the limit of high energy, the individual
components of this vector become extremely large, sufficiently so,
as we will see below, to cause scattering amplitudes to violate
unitarity.  What controls the growth of these new amplitudes?

  I will now argue that underlying local gauge invariance which is
present in the Higgs mechanism supplies the answers to both of these
questions.  To make the connection, we need one further ingredient,
which is, however, a consequence of local gauge invariance.  In any
local
field theory  in which a continuous symmetry is spontaneously broken,
the  theory  must contain a massless particle, called a {\it Goldstone
boson}.  As an example of this general principle, we might consider the
scalar field in \HiggsL.  In the above discussion, we assumed that
it is energetically favorable for $\varphi$ to acquire a vacuum
expectation value \varphivacev.  Since the theory is symmetric
under rotation of the phase of $\varphi$, this expectation value could
equally well be generated with any phase.  But now consider a field
configuration such that
$$ \VEV{\varphi(x)}  =   e^{i\alpha(x)} {1\over\sqrt{2}} v \ .
\eqn\spacevarphi$$
The phase variation shown here could at worst cost an energy proportional
 to $|\vec\nabla \alpha|$ which vanishes in the long wavelength limit.
 Thus, this phase variation  corresponds to a massless field, or, after
quantization, a massless particle.  In the following discussion, I will
denote this  particle by
$$\pi = \sqrt{2}\, {\rm Im} \varphi \ .
\eqn\pidefin$$
Notice that in a gauge theory, \spacevarphi\ is a local
 gauge transformation,
and thus the field $\pi$ can be transformed away.  Nevertheless, we
must retain it in our covariant-gauge formalism.

   An important property of a Goldstone boson is that it is created and
destroyed singly by the symmetry current.  In the example of $\varphi$,
the electromagnetic current is
$$  j^\mu = -i
 \bigl( \varphi^\dagger \d^\mu \varphi - \d^\mu \varphi^\dagger \varphi
       \bigr) \ .
\eqn\thecurrentofvar$$
Inserting \varphivacev\ and \pidefin\ into \thecurrentofvar\ to determine
the piece depending on one quantum field, we find
$$  j^\mu = v \d^\mu \pi  + \cdots
\eqn\thejdecomp$$
Then the current can create and destroy single quanta of $\pi$.  The
standard form for this relation is
$$     \bra{0} j^\mu \ket{\pi(p)} = -i F p^\mu \ ;
\eqn\standardF$$
using \thejdecomp, we can identify $F = v$ in this example.  Though
the symmetry associated with $j^\mu$ is spontaneously broken, the
current should still satisfy the equation of motion $\d_\mu j^\mu = 0$.
Applied to \standardF, this equation implies $p^2 = 0$, which confirms
that the
Goldstone boson should be massless.

\FIG\VBSELF{Vector boson self-energy in the Higgs model \HiggsL.}
  The presence in the theory of a Goldstone boson allows us to understand
 how the spin 1 particle can acquire mass compatible with current
conservation.  The structure of the vector boson self-energy in the
theory \HiggsL \ is shown in Fig. \VBSELF.  This amplitude is actually
an expectation value of two currents; thus, it should satisfy
$$     q_\mu \VEV{j^\mu(q) j^\nu(-q)} = 0 \ .
\eqn\jjiscons$$
The mass term in the Lagrangian, eq. \theHmass, contributes the term
$$           -i g^\mn  m^2  \ ,
\eqn\thehardmass$$
with $m = gv$, which does not by itself satisfy \jjiscons.  However,
because the current $j^\mu$ can create a single Goldstone boson, there
is another contribution of the same order, as shown in the figure.
This new contribution uses the matrix element \standardF\ and
contains a Goldstone boson propagator $i/q^2$.  The sum of these
contributions is
$$ \eqalign{
 -i g^\mn m^2 + (-igF q^\mu) {i\over q^2 }& (igF q^\nu) \cr
  & = -im^2 \bigl(g^\mn - {q^\mu q^\nu\over q^2}\bigr) \ , \cr}
\eqn\twotogether$$
where $m = gv = gF$.  This full expression satisfies \jjiscons.
One may, in fact, turn this argument around to show that the relation
$$              m = gF
\eqn\misgF$$
follows from the formula \standardF, independently of the underlying
Lagrangian.

 Now we have all of the ingredients we need to analyze vector boson
emission in a theory with the Higgs mechanism.  To begin, we should
write the analogue of eq. \sumofM\ for the theory with massive
spin 1 bosons.  Let $\e^\mu_{Ti}$ be the polarization vectors
corresponding to transverse polarizations, let $\e^\mu_\long$ be the
polarization vector \evector\ corresponding to longitudinal
polarization, and let $\e^\mu_t$ be a vector equal to $(1,0,0,0)$ in
the rest frame which corresponds to time-like polarization.  Then the
probability of emitting a spin 1 boson is proportional to
$$  |\e_{T1}\cdot \M|^2 + |\e_{T2}\cdot\M|^2
+ |\e_\long\cdot \M|^2 - |\e_t\cdot\M|^2 + |\M_\pi| \ .
\eqn\sumofMtwo$$
The last term in the sum involves $\M_\pi$, the matrix element for
producing a Goldstone boson.  Among these five states, the first three
are expected to be physical particles. The timelike vector boson is a
state of negative norm and must therefore be unphysical.  The Goldstone
boson is also expected to be unphysical, as explained below eq.
\spacevarphi.

\FIG\ExposeB{Emission of a single massive vector boson.}
   A relation between the latter two production amplitudes is given by
the equation of current conservation.  As in Fig.  \OnePhot, we can
analyze a vector boson emission amplitude by pulling on the vector
boson line and revealing the current to which the boson attaches.  In
the case of a massive boson, the result of that manipulation is shown
in Fig. \ExposeB.  The current can either couple directly into the
emission process, or it can couple to a single Goldstone boson which
in turn joins onto the emission diagram.  Thus, we find
$$ \VEV{j^\mu(q)} = \M^\mu  - i g F q^\mu {i\over q^2} \M_\pi \ .
\eqn\thejdec$$
If the current must be conserved, we must find zero when we contract
$q^\mu$ with this expression.  This gives the relation
$$       q_\mu \M^\mu  +    gF \M_\pi = 0 \ .
\eqn\qcontractm$$
Since $\e_t^\mu = q^\mu/m$, we find
$$   |\e_t \cdot \M |^2 = |\M_\pi|^2 \ .
\eqn\timecan$$
Thus, also in the case of a massive vector boson, the underlying
principles of gauge symmetry and current conservation guide the
cancellation of unphysical positive and negative norm states.

\FIG\GBET{The Goldstone Boson Equivalence Theorem.}
\REF\Gbett{J. M Cornwall, D. N. Levin, and G. Tiktopoulos,
            \sl Phys. Rev. \bf D10, \rm 1145 (1974).}
\REF\Gbettwo{C. E. Vayonakis, \sl Lett. Nuov. Cim. \bf 17, \rm 383
                    (1976).}

 This argument can be pushed a bit farther to develop an additional
   piece of
insight.  Notice that the longitudinal polarization vector \evector\
satisfies
$$  \e^\mu_\long   = {q^\mu\over m} + {\cal O}\bigl({m\over q}\bigr)\ .
\eqn\limitoflong$$
I  have already remarked that the individual components of $\e^\mu_\long$
can be extremely large.  However, we now see that
$$ \e_\long \cdot \M \simeq \e_t \cdot \M  \ ,
\eqn\epslongM$$
which is in turn related by \timecan\ to the amplitude for emission of a
Goldstone boson.  Thus, we find the relation shown in Fig. \GBET,
known as the {\it Goldstone Boson Equivalence
Theorem}.\refmark{\Gbett,\Gbettwo}
This formula has the following physical interpretation:  Through the
Higgs mechanism, the vector field becomes massive by eating the
the Goldstone boson.  At high energy, the spontaneous symmetry breaking
becomes irrelevant, and the emission amplitudes for massive bosons
show their origin as a combination of transverse spin 1 and Goldstone
boson emission amplitudes.

\REF\mgk{M. S. Chanowitz and  M. K. Gaillard,
          \sl  Nucl. Phys. \bf B261, \rm 379 (1985).}
\REF\bs{J. Bagger and C. Schmidt, \sl Phys. Rev. \bf D41, \rm 264
                      (1990).}
\REF\velt{H. Veltman, \sl Phys. Rev. \bf D41, \rm 2294 (1990).}
   The argument for the Goldstone Boson Equivalence Theorem is given here
only at the simplest level.
A more careful argument is needed when several vector bosons are emitted
and when loop corrections to the boson propagators are included.
However, the theorem remains true in these situations.  Some recent
analyses which take account of these subtleties are given in refs.
\mgk--\velt.

\section{Examples from the Standard Model}

If the discussion of the previous section was a bit abstract, the
moral of this discussion has direct application to high energy processes
in the standard electroweak gauge theory.  In this section, I will
present two important examples.

The first of these is the theory of the top quark width.  We now know
that the top quark is sufficiently heavy to decay to an on-shell $W$
boson and a bottom quark.  For this two-body
decay, one might roughly estimate the width as  $\Gamma_t \sim (\alpha_w/
4\pi) m_t$, where $\alpha_w =  g^2/4\pi = \alpha/\sstw$.
  The width of the top quark eventually controls
the qualitative features of top decays, so it is important to understand
its magnitude.  Surprisingly, this rough estimate turns out to be very
naive; the true result for $\Gamma_t$ grows as $m_t^3$.   The explanation
for this change comes from the Goldstone Boson Equivalence Theorem.

\FIG\TopWidth{Diagrams contributing to the top quark width: (a) the
          leading order contribution in the standard model; (b) the
           analogous Goldstone boson diagram.}
   Before I explain the behavior of the top quark width, let us obtain
the correct result by a straightforward Feynman diagram calculation.
In the standard model, the top quark width is given to leading order
by the diagram of Fig. \TopWidth(a).  The decay matrix element is
$$ i \M = {ig\over \sqrt{2}} \bar u(p_b) \gamma^\mu {(1-\gamma^5\over 2}
                   u(p_t)  \, \e_\mu^*(q) \ ,
\eqn\topmat$$
where $q$ is the $W$ momentum.  From here on, ignore the $b$ quark mass.
Then the square of the matrix element is
$$ \half \sum_{\rm spins} |\M|^2 = {g^2\over 2}\bigl[p_b^\mu p_t^\nu
     + p_b^\nu p_t^\mu - g^\mn p_b\cdot p_t\bigr] \e^*_\mu(q) \e_\nu(q)
                  \ .
\eqn\sqtopmat$$
Now sum over the three physical $W$ polarizations, excluding the timelike
polarization:
$$   \sum_{\rm pol}  \e^*_\mu(q) \e_\nu(q)
        = - \bigl(g_\mn - {q_\mu q_\nu\over m_W^2}\bigr) \ .
\eqn\poolsum$$
This gives
$$ \half \sum_{\rm spins} |\M|^2 = {g^2\over 2}\bigl[p_b \cdot p_t
    + 2 {q\cdot p_b q\cdot p_t\over m_W^2} \bigr]
                  \ .
\eqn\sqtopmattwo$$
To simplify this expression, use the kinematic relations
$$2p_b\cdot p_t = 2p_b\cdot q =
m_t^2 - m_W^2\ , \qquad 2p_t \cdot q = m_t^2 + m_W^2 \ .
\eqn\kinem$$
Notice that the second term in \sqtopmattwo\ is of order $(m_t^4/m_W^2)$.
Add phase space factors  to obtain the final result
$$  \Gamma_t = {g^2\over 64\pi} {m_t^3\over m_W^2}
          \bigl(1 - {m_W^2\over m_t^2}\bigr)^2
          \bigl(1 + 2 {m_W^2\over m_t^2}\bigr)
\eqn\fullwidth$$
As promised, this result grows as $m_t^3$.

  The result we have found is doubly surprising because the large result
comes from the $q_\mu q_\nu$ terms in \poolsum.  In quantum
electrodynamics, this term in the spin sum always cancels out due to
current conservation. But in weak interaction theory, the current
$\bar b \gamma^\mu (1-\gamma^5) t$ which mediates the top decay is not
conserved; in fact, its divergence is of order $m_t$.

  However, in the context of our discussion of the interplay of gauge
symmetry and Goldstone bosons, the result is easily understood.  Though
the quark charged current is not conserved, one can add terms involving
Goldstone bosons to form the conserved gauge current of a spontaneously
broken gauge theory.  The analysis of the previous section applies to
this theory directly.  Thus, the Goldstone Boson Equivalence Theorem
tells us that the leading behavior of the top quark width at high
energy should be given by the Goldstone boson emission diagram of
Fig. \TopWidth(b).  I will now compute this diagram and verify the
correspondence.

  The matrix element for the emission of a Goldstone boson from a
top quark, as shown in Fig. \TopWidth(b), is
$$ i \M =  - \lambda_t \bar u(p_b){(1+\gamma^5\over 2}
                   u(p_t)   \  .
\eqn\topmatG$$
In this expression, $\lambda_t$ is the coupling of the top quark to the
Higgs boson.  In the standard model,
both the top quark mass and the $W$ boson mass arise
from the Higgs field vacuum expectation value $v$, according to the
relations
$$   m_t = {\lambda_t v\over \sqrt{2}} \ , \qquad
   m_W = {g v\over  2} \ .
\eqn\topandWmass$$
Using these formulae to eliminate $\lambda_t$ and $v$, we find
$$ \half \sum_{\rm spins} |\M|^2 = \lambda_t^2 p_b\cdot p_t =
{g^2 m_t^2\over 2 m_W^2}\cdot {m_t^2\over 2}
\eqn\sqtopmatG$$
This leads to an expression for the top quark width
$$  \Gamma_t = {g^2\over 64\pi} {m_t^3\over m_W^2}
\eqn\fullwidthG$$
which does indeed capture the leading behavior of \fullwidth.

   \REF\Alles{W. Alles, C. Boyer, and A. J. Buras,
         \sl Nucl. Phys. \bf B119, \rm 125 (1977).}
 A more complex process which is strongly affected by the physics of
massive spin 1 particles is the reaction $\ee\rarr \pairof{W}$.
I will present a semiquantitative discussion of this reaction; the
full tree-level formulae can be found, for example, in ref. \Alles.

\FIG\WW{Diagrams contributing to $\ee\rarr \pairof{W}$: (a) the
          leading order contributions in the standard model; (b) the
           analogous Goldstone boson diagrams.}
   The leading order diagrams contributing to $\ee\rarr \pairof{W}$
in the standard electroweak model are shown in Fig. \WW(a).  To
understand the conceptual issues which this process addresses, let us
make a naive estimate of the first of these diagrams.  Roughly,
we expect the amplitude for $W$ pair production by a virtual photon
to be given by the amplitude for production of charged scalars, times
a polarization inner product:
$$   i\M_\gamma \sim  i\M(\ee \rarr \phi^+\phi^-) \cdot \e^{*\mu}(q_+)
              \e^*_\mu(q_-) \ .
\eqn\firsttryMee$$
However, for the case of longitudinally polarized $W$ bosons, the
polarization product in \firsttryMee\ is very large.  We can apply
\limitoflong\ to estimate
$$\e^{*\mu}(q_+)    \e^*_\mu(q_-) \simeq {q_+\cdot q_-\over m_W^2} =
              {s\over 2 m_W^2} \ .
\eqn\epsprod$$
This estimate of the matrix element of $W$ pair production would
imply
$$    {d\sigma\over d \cos\theta} \sim {\pi\alpha\over s} \cdot
      \bigl( {s\over 2 m_W^2} \bigr)^2 \ .
\eqn\firsttrysig$$
But this result is unphysically large.  The $(1/s)$ behavior of the
first factor in \firsttrysig\ is actually the largest asymptotic
behavior allowed by unitarity for a single partial wave.  Somehow, the
strong energy dependence of \epsprod \ must be cancelled in the
full result for the cross section.

  Our general analysis of massive vector bosons tells us that this
cancellation must occur, and that the matrix element for longitudinal
$W$ pair production must eventually be reduced to that for Goldstone
boson pair production.  I will sketch how this works in the amplitude
for annihilation of polarized electrons, $\M(e^-_L e^+_R \rarr W^+_\long
 W^-_\long)$.

 The first two diagrams in Fig. \WW(a) have the same
general form, in which the two $W$ polarization vectors dot into the
Yang-Mills vertex function.  Using the approximation \limitoflong\
for the longitudinal polarization vectors, one can simplify this
vertex and arrive at the following expression:
$$ \eqalign{
i\M_{\gamma+Z} = -ie^2 (q_+-q_-)^\mu  \bar v(p_+) &\gamma_\mu
 {(1-\gamma^5)\over 2} u(p_-) \crr &
 \cdot \bigl({1\over s} + {\half - \sstw
  \over \sstw}{1\over s- m_Z^2} \bigr) \cdot {s-2m_w^2\over 2 m_W^2} \ .
\cr}
\eqn\gamZmats$$
In this expression, the first term is the matrix element for scalar
boson production, the term in parentheses is the coherent sum of
virtual photon and $Z$ propagators, and the final term is the
result of contracting longitudinal polarization vectors with the
three-boson vertex.  Though there is some cancellation between the
photon and $Z$ contributions, the full result still shows the
pathology described in the previous paragraph.

 The third diagram of Fig. \WW(a) has a different kinematic structure.
However, when one contracts this diagram
with the longitudinal polarization vectors, one finds
$$ i\M_\nu = -i{e^2\over 2 \sstw}  \bar v(p_+) {\gamma\cdot q_+\over m_W}
{\gamma \cdot (p_- - q_-)\over (p_- - q_-)^2} {\gamma\cdot q_-\over m_W}
       {(1-\gamma^5)\over 2} u(p_-) \ .
\eqn\thenuterm$$
  Since $p_- u(p_-) = 0$, we can replace $(\gamma\cdot q_-)$ by
$\gamma \cdot (q_- - p_-)$; this factor cancels the neutrino propagator.
Then the expression \thenuterm\ can be rearranged into
$$ i\M_\nu = -ie^2 (q_+-q_-)^\mu \bar v(p_+)\gamma_\mu
       {(1-\gamma^5)\over 2} u(p_-) \cdot  \bigl( - {1\over 2\sstw}
                        {1\over 2m_W}\bigr)
\eqn\thenutermtwo$$

Now add \gamZmats\ and \thenutermtwo, take the high energy limit, and
use the standard model relation $m_W^2 = m_Z^2 \cstw$.  This gives
$$\eqalign{
 i\M & \simeq -ie^2 (q_+-q_-)^\mu \bar v(p_+)\gamma_\mu
       {(1-\gamma^5)\over 2} u(p_-) \crr & \hskip 1in
        \cdot  \bigl( {m_Z^2\over s -m_Z^2}
              {\half - \sstw\over 2 m_W^2 \sstw}- {1\over s-M_Z^2}
                     {1\over 2 \sstw}\bigr) \crr
 & \simeq -ie^2 (q_+-q_-)^\mu \bar v(p_+)\gamma_\mu
       {(1-\gamma^5)\over 2} u(p_-) \cdot {1\over s} \cdot
 \bigl({1 - 2 \sstw - 2 \cstw \over 4\sstw\cstw} \bigr) \crr
 & \simeq +ie^2 (q_+-q_-)^\mu \bar v(p_+)\gamma_\mu
       {(1-\gamma^5)\over 2} u(p_-) \cdot {1\over s} \cdot
 \bigl({1\over 4\sstw} + {1\over 4 \cstw}\bigr) \ . \cr}
\eqn\Mrarrange$$
This last answer is exactly the result of computing the diagrams of
Fig. \WW(b), in which the gauge bosons of the
standard weak interaction model create pairs of Goldstone bosons.
In the expression in parentheses in the last line, the first
term, with the coupling $(e/\sin\theta_w)$, is the contribution of
$SU(2)$ boson exchange, while the second term, with the coupling
$(e/\cos\theta_w)$, is the contribution of $U(1)$ boson exchange.
The final answer not only respects unitarity but actually is smaller
than the amplitude for the pair production of transversely polarized
$W$ bosons.  The cancellations that lead to this point are organized
by the Goldstone Boson Equivalence Theorem and the underlying principle
of exact local gauge invariance.

\chapter{Higher Spin}

 We have now seen that the theory of spin 1 fields and their associated
particles is surprisingly complex.  In particular, it requires a higher
principle such as current conservation to organize the states created
by the field and to neatly cancel all contributions except those from
physical propagating modes.  These cancellations must occur even more
strongly and more intricately in theories of spin greater than 1.
I will now explain how our earlier arguments generalize to these cases.
In this discussion, I will concentrate on theories of massless particles.
As was demonstrated in the previous section, the corresponding massive
theories are built from the massless theories and are, if anything, more
highly constrained.

 In my general discussion of the connection between fields and
 particles, I pointed out that, as the spin of a field increases,
fewer and fewer of its components create and destroy
physical propagating states.
 In general, only the components of
maximal helicity are physical.  Thus, from the 8 complex-valued
components of a spin $\thalf$ field, only two---$\psi_{+ \dot + \dot +}$
and $\psi_{-\dot- \dot -}$---create and destroy physical particles
moving in the $\hat 3$ direction.  All other states which are created
by the various components of
$\psi_{\a \dot a \dot \b}$ in some mathematical formalism must be
made to cancel out.  This applies most strongly to the states of
negative norm created by $\sigma^{0\a\dot a}\psi_{\a\dot a\dot b}$.

  The cancellation of these unphysical components occurs
naturally, just as in the spin 1 case, when the higher spin field
couples to a conserved tensor.  Thus, we can make a consistent
theory of a massless spin $\thalf$ field $\psi_{\mu \dot \a}$
in a theory which contains a conserved spin $\thalf$ tensor
current $s_{\mu \dot\a}$, with the cancellations of unphysical
modes following from the pair of equations
$$         \d^\mu   s_{\mu\dot\a} = 0 \ .
\eqn\Scons$$
Similarly, we can construct a consistent theory of a massless
spin 2 field $h_{\mu\nu}$ by coupling it to a conserved
two-index tensor $t_{\mu\nu}$.  In Yang-Mills theory, the coupling
to the gauge field changes the current conservation
equation $\d^\mu j_\mu = 0$ to a modified, gauge-covariant equation
 $D^\mu j_\mu = 0$ which
agrees with the standard equation to leading order.
Such a modification is also typical in theories of higher spin.

\REF\Wgen{S. Weinberg, \sl Phys. Rev. \bf 138, \rm B988 (1965).}
\REF\Blgen{D. G. Boulware and S. Deser, \sl Ann. Phys. \bf 89, \rm
                 193 (1975).}

  To construct a theory of higher spin fields, we must thus ask,
what conserved tensors of higher spin are available to be the sources
of the new fields?  One candidate is obvious.  The energy-momentum
tensor of all particles and fields, $T_\mn$, is naturally conserved
and can be considered as the source of a spin 2 field.  The gauge
theory of spin 2 which results from this coupling is precisely
general relativity.  The conservation law of the energy-momentum
tensor is modified self-consistently to
$$           \nabla^\mu  T_{\mn} = 0 \ ,
\eqn\newconsT$$
the analogue of the covariant conservation law for the Yang-Mills
current.  General relativity contains only helicity $\pm 2$ particles
as propagating states, and the production of these particles
from the conserved energy-momentum tensor naturally cancels
additional, unphysical modes.\refmark{\Wgen,\Blgen}

\REF\CMan{S. Coleman and J. Mandula, \sl Phys. Rev. \bf 159,
           \rm 1251 (1967).}
   In order to construct a higher-spin field in addition to the
gravitational field, we must identity a second naturally conserved
tensor.  Unfortunately, this is extremely difficult; almost every
possible case is excluded by general restrictions on the $S$-matrix
proved by Coleman and Mandula.\refmark{\CMan}
To understand the origin of these restrictions, let us consider the
constraints on the existence of a second conserved two-index tensor
$r_{\mn}$, in addition to the full energy-momentum tensor $T_\mn$.

  The spatial integrals of $T_\mn$ give a globally conserved
  energy-momentum 4-vector $P^\mu$.  Similarly, let us define the
4-vector
$$          R^\mu = \int d^3 x \, r^{0\mu} \ .
\eqn\Rdefin$$
The vector $R^\mu$ is an additional conserved quantity which
restricts scattering processes.  By Lorentz covariance, the  diagonal
matrix elements of $R^\mu$ in one particle states of momentum $p$
are proportional to $p^\mu$,
$$  \bra{p,a}  R^\mu \ket{p,a} =  C_a  p^\mu \ ,
\eqn\rexpval$$
where the constant of proportionality $C_a$ depends only on the
particle type.  Now consider the elastic scattering of particles
of two different types, $1 + 2 \rarr 1 + 2$.  Conservation of
$P^\mu$ yields the constraint
$$   p_1^\mu + p_2^\mu = p_1^{\prime \mu} + p_2^{\prime \mu} \ .
\eqn\psum$$
Conservation of $R^\mu$ yields the additional equation
$$  C_1 p_1^\mu + C_2 p_2^\mu = C_1
p_1^{\prime \mu} + C_2 p_2^{\prime \mu} \ .
\eqn\rsum$$
The constraint \psum\ is solved by going to the center of mass frame;
  \FIG\Coleman{Constraints on elastic scattering imposed by conservation
       of two 4-vectors $P^\mu$ and $R^\mu$.}
then (if the final state remains in the $\hat 1$--$\hat 3$ plane)
the allowed values of $\vec p_1^\prime$ and $\vec p_2^\prime$
lie on a circle.  This constraint is shown in Fig. \Coleman.
In this frame, the constraint \rsum\ restricts the vector $\vec
p_1^\prime$  to an ellipse.  The two constraints intersect for forward
scattering and possibly at some additional specific angles.  However
the general property that the $S$-matrix is analytic in the momentum
transfer $t$ forbids such discrete solutions.  We conclude that, if
both conditions \psum\ and \rsum\ are to be imposed simultaneously,
there can be no elastic scattering of 1 from 2.

  The theorem of Coleman and Mandula\refmark{\CMan}  generalizes this
argument to forbid additional conserved 4-vectors and conserved
tensors of any higher rank (except for the Lorentz group
generators $M_\mn$).
 Thus, it implies that gravity is the
only consistent theory of a spin 2 field, and that there are no
consistent theories of massless fields with spin higher than 2. At least,
no such theory can be constructed according to the strategy described
here.

The case of spin $\thalf$ is more ambiguous. A conserved spin $\thalf$
current $s_{\mu\dot a}$ leads to a conserved spinor charge $Q_{\dot a}$.
However, such a charge does not have diagonal matrix elements in
single particle states.  Thus, it is not necessarily forbidden, but
its properties are strongly restricted. We will discuss this case in
Section 7.  For the  case of spin ${5\over 2}$, the loophole available
for spin $\thalf$ can be closed, and the required source is
forbidden by the Coleman-Mandula theorem.  Thus, we come to the end of
our catalogue of possible higher-spin fields.

\chapter{Spin $\half$ as a Construct}

 Now I will turn to the question with which we began
these lectures:  Why is there spin $\half$?  In this section and the
next, I will describe two possible solutions.  In this section, I will
show how to construct spin $\half$ from a mechanical model of particle
dynamics.  In the next section, I will describe a symmetry, which might
be a fundamental symmetry of Nature, which requires spin $\half$ to
complete its symmetry multiplets.

\section{A Model of a Scalar Particle}

  Before we can form a mechanical model which produces relativistic
spin $\half$ particles, we should construct a model which leads to
ordinary relativistic scalar particles.  This is easily done by
imagining scalar particles as point objects which move through
space-time along world lines, and then write the mathematics
appropriate to this physical picture.  A natural guess is that the
quantum mechanical amplitude for a particle to propagate from
the space-time point $y$ to $x$ is given by an integral over paths
$$    D(x,y) = \int \D X^\mu  \, \exp\bigl( i S[X^\mu] \bigr) \ ,
\eqn\beginD$$
where $X^\mu(s)$ is a path from $y$ to $x$
$S[X]$ is some appropriate phase that the particle's wave function
acquires as it moves along the path.  A particle of mass $m$ at rest
would be expected to acquire a factor
$$                e^{-im t} \ ;
\eqn\themassphase$$
thus, a reasonable guess for $S$ is that it is proportional to the
proper time which elapses along the path:
$$          S = - m \int ds\,\sqrt{ \bigl({dX^\mu \over d s}\bigr)^{2} }
                          \ .
\eqn\theSofproptime$$

 Does the expression \beginD\ with \theSofproptime\ really lead to a
description of relativistic scalar particles?  It will be more
straightforward to work with this expression if we rewrite it in
such a way that the square root in
\theSofproptime\ is removed.  To do this,
introduce a new parameter $e(s)$
which is a function of the position on the path, and write
$$    D(x,y) = \int \D X^\mu \D e \, \exp\Bigl( -i
  \int ds\, \half \bigl[{ (\dot X)^2\over e} + em^2 \bigr] \Bigr)\ ,
\eqn\secondD$$
where $\dot X^\mu = dX^\mu/ds$.  The variational equation for $e(s)$
is
$$           -{(\dot X)^2\over e^2}  + m^2 = 0 \ ;
\eqn\determineE$$
fixing $e(s)$ as the solution to this equation, and substituting into
the exponent of \secondD,    we recover
\theSofproptime.  Thus, \secondD \ is also a reasonable starting point
for our discussion.

In the construction of \theSofproptime\ and \secondD,  position along
the path is parametrized by the coordinate $s$.  However, in both
expressions for the path integral,
the choice
of the parameter $s$ is arbitrary.  Both exponents have the local
invariance
$$    X^\mu(s) \rarr X^\mu(g(s)) \ ,\qquad e(s) \rarr {dg\over ds}\,
 e(g(s)) \ ,
\eqn\reparam$$
corresponding to the change of variables $s\rarr g(s)$.  Since this
is an invariance at each point $s$, it is a gauge symmetry of the
path integrals, and these integrals must be defined by Fadde'ev-Popov
gauge fixing.  The gauge freedom of \secondD\ can be fixed in a
simple way:  Set
$$             e(s) = 1         \ .
\eqn\fixedgauge$$
There are no ghosts or other awkward consequences of this choice of
gauge.

With this prescription, the general
sum over paths can be written as a sum over paths for which the
parameter $s$ runs from 0 to $T$, and an integral over $T$. We thus
obtain
$$ D(x,y) = \int_0^\infty dT \, \int \D X^\mu \exp\Bigl( - i
  \int ds\, \half \bigl[ (\dot X)^2 + m^2 \bigr] \Bigr)\ ,
\eqn\thirdD$$
This functional integral can be evaluated explicitly.  It is, in fact,
the Feynman path integral for a nonrelativistic system  with
Hamiltonian
$$H = m^2 - p^2 = m^2 + (\vec p)^2 - (p^0)^2  \ ,
\eqn\thepHam$$
 integrated
over time $T$:
$$ \eqalign{
D(x,y) &= \int_0^\infty dT \, \bra{x} e^{-i(m^2-p^2)T}\ket{y}\cr
       &= { i \over p^2 - m^2}     \ . \cr}
\eqn\finalD$$
The final result is the Feynman propagator for a free scalar particle,
the best result we could have hoped for.

Though
we set up this construction by considering the scalar particle as an
 world line embedded in space-time, an alternative viewpoint
is possible.  We could as well interpret eq. \secondD\ or \thirdD\ as
representing  an abstract world line, with a one-dimensional
quantum field theory living on it.  The Lagrangian of this quantum
field theory is
$$      L  =  - \half \bigl[ (\dot X)^2 + m^2\bigr] \ .
\eqn\ellofone$$
In this view, the space-time coordinates $X^\mu(s)$ are fields which
are a part of this one-dimensional field theory.  In other words,
one may view space-time as living {\it on}
 the world line as easily as one
might view the world line as living {\it in} space-time.

\section{Addition of Spin}

\REF\Ramond{P. Ramond, \sl Phys. Rev. \bf D3, \rm 2415 (1971).}
\REF\BDVH{L. Brink, P. Di Vecchia, and P. Howe,
        \sl Nucl. Phys. \bf B118, \rm 76 (1977).}
  From the point of view in which space-time is an attribute of the
particle's world line, it is easy to find generalizations which
produce more interesting types of particles.  A simple way to
modify \ellofone\ is to add an anticommuting coordinate field
$\theta^\mu(s)$, $\mu = 0,1,2,3$.\refmark{\Ramond,\BDVH}
  The Lagrangian of this extended
theory is
$$      L  =  - \half \bigl[ (\dot X)^2 + \theta^\mu \dot\theta_\mu
                    + m^2 \bigr] \ .
\eqn\elloftwo$$

In principle, we might have tried adding any field that could live
on the 1-dimensional world line.  This particular choice, however,
is especially interesting because the Lagrangian \elloftwo\ has an
unusual symmetry.  Consider the transformation generated by
$$   \delta X^\mu = \e \theta^\mu \ , \qquad
   \delta \theta^\mu =  - \e \dot X^\mu \ ,
\eqn\newtranf$$
where $\e$ is an anticommuting number:  $\e_1\e_2 = - \e_2\e_1$.
The second variation under this transformation is
$$  ( \delta_1 \delta_2  - \delta_2 \delta_1) A = 2\e_2\e_1 \dot A\ ,
\eqn\Avar$$
where $A$ is $X^\mu$ or $\theta^\mu$.  Thus, the transformation
\newtranf\ is in some sense the square root of a translation in $s$.
It is not difficult to show that
\newtranf\ is a symmetry of the Lagrangian \elloftwo:
$$\delta \int ds L = \int ds \bigl( \dot X^\mu \e \dot\theta_\mu
             - \e \dot X^\mu \dot\theta_\mu \bigr) = 0 \ .
\eqn\newsymm$$
For the moment, we might view this symmetry as an amusing feature of
this particular extension; we will have more to say about it in
Section 7.

  In order to find the interpretation of \elloftwo\ in terms of
particles, we need to find the analogue of eq. \finalD\ for
the  one-dimensional functional integral which contains this
Lagrangian.  The part of \elloftwo\  which contains commuting
numbers is treated just as in \finalD.  For the anticommuting
numbers, the Lagrangian term $\half\theta\cdot \dot\theta$ should be
compared to the standard expression for commuting numbers
$$         L = \sum_i p_i \dot q_i  - H(p,q) \ .
\eqn\genformL$$
To make the analogy, we  take the half of the
$\theta^\mu$ to be
 canonical coordinates and the other half to be  their conjugate
momenta, with $H=0$.  Since these objects anticommute, we should convert
them to quantum operators
 with anticommutation relations.  Then the
quantum operators $\theta^\mu$  obey
$$          \big\{\theta^\mu , \theta^\nu \big\} = - \half g^\mn \ .
\eqn\theanticomm$$
One way to interpret these set of relations is to diagonalize it
by finding two linear combinations of the $\theta^\mu$, which might be
called $a_i$, and two orthogonal linear combinations
$a^\dagger_i$ which obey
$$          \big\{  a_i, a^\dagger_j \big\} = \delta_{ij} \ .
\eqn\acomms$$
Then the Hilbert space acted on by the $\theta^\mu$ is described by
four states
$$\ket{\phi}\ ,\quad a^\dagger_1 \ket{\phi}
\ ,\quad a^\dagger_2 \ket{\phi}
\ ,\quad a^\dagger_1  a^\dagger_2 \ket{\phi} \ .
\eqn\fourphistates$$
Alternatively, we might recognize that the algebra \theanticomm\ is
exactly the Dirac algebra  \Diracalg.  Then the four states
indicated schematically in \fourphistates \ correspond to the
four-dimensional Dirac representation of the Lorentz group.  Thus, the
particle which moves along the world line carries
a Dirac spinor and thus has spin $\half$.

This particular constructive picture of spin works uniquely for spin
$\half$.  However, it is easily generalized to provide a construction
of spin 1, and also higher spins.  To construct particles with spin 1,
choose the Lagrangian
$$      L  =  - \half \bigl[ (\dot X)^2 + m^2 \bigr] - \bar\theta^\mu
         \dot \theta_\mu + \omega \bar\theta^\mu \theta_\mu  \ ,
\eqn\ellofthree$$
where now $\theta^\mu$ is a complex anticommuting number, with
$\bar\theta^\mu$ its complex conjugate.  Interpreting these
variables as canonical coordinates and momenta according to
\genformL, we are led to the set of commutation relations
$$          \big\{\bar\theta^\mu , \theta^\nu \big\} = - g^\mn \ .
\eqn\theanticommtwo$$
and the Hamiltonian
$$      H = - \omega  \bar\theta \cdot \theta \ .
\eqn\thenewHam$$
Given \theanticommtwo, it is natural to interpret $\theta^\mu$ as a
set of fermionic annihilation  operators $a^\mu$
(with positive metric for
$\mu = 1,2,3$) and $\bar\theta^\mu$  as the corresponding
creation operators $a^{\dagger\mu}$.  The Hamiltonian is
$H = \omega a^\dagger\cdot a$.  Then this theory contains
particle world lines associated with the various possible fermion
states:
$$ \eqalign{
  \ket{0} \ & \qquad {\rm spin} \ 0\ , \quad ({\rm mass})^2 = m^2 \cr
a^{\dagger i}
  \ket{0} \ & \qquad {\rm spin} \ 1\ , \quad ({\rm mass})^2 = m^2 +
                     \omega \ ,  \cr }
\eqn\manystates$$
and so on.

  Though this analysis does give a construction of spin 1, it raises
as many questions as it solves.  For example, what kind of field
is associated with $a^{\dagger i}a^{\dagger j}\ket{0}$?  This
field should be present in the theory  unless it is removed by
taking $\omega$ very large. There are problems in
obtaining {\it massless} spin 1 particles:  If $(m^2 +\omega)=0$,
then either scalar particles or the new states with two world-line
fermions will have
negative (mass)${}^2$.  Finally, the construction contains explicit
negative norm states $a^{\dagger 0}\ket{0}$.  These states
must have a mechanism to cancel completely from all scattering
processes.  It is possible to give satisfactory answers to these
questions, but only within a formal structure more constraining
than the particle models discussed in this section.  We will find a
better setting for addressing these questions in Section 8.

\chapter{Spin $\half$ as a Symmetry}

  As an alternative to explaining spin as a mechanical attribute of
particles, one might attempt to postulate a symmetry of Nature which
naturally leads to spin $\half$.  In this type of model, one would
postulate that there exists  an operator $Q$ which generates a
symmetry of the Hamiltonian of particle interactions and also has
the property of converting particles of zero spin into particles with
spin $\half$.  Such a symmetry is known as a `supersymmetry'.
The first renormalizable field theory with a supersymmetry was
constructed by
Wess and Zumino.\Ref\WandZ{J. Wess and B. Zumino, \sl Nucl. Phys.
       \bf B70, \rm 39 (974).}    The formal
       consequences of supersymmetry
are presented in detail in the book of Wess and Bagger.\Ref\WBbook{J.
       Wess and J. Bagger, \sl  Supersymmetry and Supergravity. \rm
       (Princeton University  Press, 1983).}

 \section{The Supersymmetry Algebra}

 Any charge which converts spin 0 to spin $\half$ must
carry  a spinor index.  The simplest choice, and actually the
only consistent one, is to take this charge to carry spin $\half$.
However, the hypothesis of a spin $\half$ charge $Q_\alpha$ which
commutes with the Hamiltonian turns out to be extremely restrictive.
To see this, write the anticommutator of the charge $Q_\a$
with its Hermitian conjugate:
$$             \big\{ Q_\a, Q^\dagger_{\dot \a} \big\} = R_{a\dot\a}
\eqn\firstsusycomm$$
If $Q$ commutes   with the Hamiltonian, $Q^\dagger$ will as well;
thus $R_{\a\dot\a}$ commutes with the Hamiltonian.  We recognize
$R_{\a\dot\a}$ as a conserved vector, just the sort of object which
was excluded by the Coleman-Mandula theorem, as described in
the discussion following eq. \Rdefin.  On the other hand, $R_{\a\dot\a}$
cannot be zero:  Since $R$ is the square of $Q$, $R = 0$ only if
 both $Q$ and $Q^\dagger$ give zero on all states of the Hilbert
space.

 There is only one way out of this dilemma.  $R_{\a\dot\a}$ must be
the one conserved 4-vector allowed by the Coleman-Mandula
theorem---the total energy-momentum $P^\mu$.  Thus, if $Q_\a$ is a
symmetry of the Hamiltonian carrying spin $\half$, it must
obey the commutation relation
$$   \big\{ Q_\a, Q^\dagger_{\dot \a} \big\} = 2 \sigma_{a\dot\a}^\mu
                       P_\mu \ .
\eqn\susycomm$$
No half-measures are possible.  $Q_\a$ must be a fundamental
symmetry of space-time, generalizing the Poincar\'e algebra.  The
new symmetry generated by $Q_\a$ must act on every particle in Nature.
Thus, we are led to a profound generalization of the theory of
elementary particles.

  To understand the consequences of the symmetry algebra \susycomm\
a bit better, consider the representation of this algebra in the
simplest context---massless
single-particle states moving in the $\hat 3$ direction.  For such
states, we saw that in Section 2.5 that
the Poincar\'e algebra has one-dimensional
representations, plus their reflections under $CPT$. We will now
analyze how the algebra \susycomm\ links these representations.

  The first step in working with \susycomm\ is to
write
the $+\dot +$ and $- \dot-$ components of \susycomm\ explicitly:
$$ \eqalign{
   \big\{ Q_+, Q^\dagger_{\dot +}\big\} &= 2 ( H - P^3) \cr
   \big\{ Q_-, Q^\dagger_{\dot -}\big\} &= 2 ( H + P^3) \cr  }
\eqn\susycommml$$
Since the quantities on the right-hand side generate translations
in time and space,
these commutation relations are reminiscent of \Avar.  In fact, they
represent the correct generalization of \Avar\ to a multidimensional
space-time.

   A massless particle moving in the $\hat 3$ direction
satisfies $(H-P^3) = 0$, $(H+P^3) = 2P^3$.  Thus, $Q_+$ and
$Q^\dagger_{\dot +}$ must give zero on such states, while $Q_-$
and $Q^\dagger_{\dot -}$ give a nonzero result. Define
$$  {1\over \sqrt{4P^3}} Q_- = a \ , \qquad
  {1\over \sqrt{4P^3}} Q^\dagger_{\dot -} = a^\dagger \ .
\eqn\newas$$
Then the second line of \susycommml\ becomes
$$ \big\{ a,a^\dagger\big\} = 1  \ .
\eqn\susya$$
The operators \newas\ thus act on the one-particle states as
fermion creation and annihilation operators.  Since these operators
raise and lower the $\hat 3$ component of angular momentum, and thus
the helicity, we can view the pairs of states with and without the
fermion created by $a^\dagger$ as
 pairs of states  $\ket{p,\lambda}$:
$$\eqalign{
 \ket{p,0}     &\leftrightarrow  \ket{p,\half} \cr
 \ket{p,\half}     &\leftrightarrow  \ket{p,1}\ , \cr}
\eqn\arels$$
and so forth.
These relations clarify the intuitive idea that
supersymmetry links bosonic and fermion particle states.

\section{Supersymmetric Dynamics}

 The requirement of  supersymmetry thus forces a quantum field theory
to contain spin $\half$ particles and fields.  According to
eq. \arels, and its generalization to massive states, a spin zero
particle in a supersymmetric theory must have a spin $\half$
partner, and a spin 1 particle must have either a spin $\half$ or a
spin $\thalf$ partner.  The interactions of these new fermions are
linked to the interactions of the bosons  through the constraint
of supersymmetry.

\FIG\Intss{Supersymmetry specifies the interactions of fermions from
    the interactions of the corresponding bosons, or vice versa: (a)
    the interaction vertex for the spin $\half$ partner of a gauge
   boson, (b) the interaction vertices for the spin 0 partner of the
      left-handed electron.}
 Perhaps the most interesting case from the viewpoint of first
principles is the relation between spin 1 and spin $\half$.  Local
gauge invariance requires the existence of spin 1 particles.  Since
supersymmetry in turn requires the existence of spin $\half$ particles,
it seems that we might construct a complete rationale for the
particles which compose the standard model.  However, the details do
not fall into place correctly.

  Given the existence of gauge bosons,
supersymmetry specifies the
quantum numbers and interactions of the new fermions.  In the
simplest realization of supersymmetry, the global symmetry
charges commute with the supersymmetry charges:
$$         \bigl[ Q_A,Q_\a\bigr] = 0 \ .
\eqn\AAcomm$$
We will see below that the strong constraints from this relation
are not made weaker in more complicated realizations of the algebra.
{}From \AAcomm, the partners
of a set of gauge bosons $A_\mu^A$ are fermions $\lambda_\a^A$
which belong to the same (adjoint) representation of the gauge
group.  As shown in Fig. \Intss(a), the Yang-Mills vertex for
gauge bosons induces an interaction between the gauge boson
and the new spin $\half$ particle.  This interaction is
exactly the one present in the simple Lagrangian
$$  \L = - {1\over 4} (F_\mn^A)^2 + \bar\lambda \,i \sigma\cdot D \lambda
 \ .
\eqn\newLagforlam$$
The Lagrangian \newLagforlam \ is the simple minimal coupling of a
gauge particle to a chiral fermion, but it happens that, when the
fermion and the gauge boson belong to the same representation of the
gauge group, this Lagrangian is supersymmtric.

  Unfortunately, the representation assignment of the fermions is
precisely not what is needed to construct the standard model.
{}From $W$ bosons in the $I=1$ representation of weak interaction $SU(2)$,
supersymmetry would require $I=1$ fermions $\widetilde w_\a$.  From
the gluons, which belong to the octet  representation of color $SU(3)$,
supersymmetry would require a  multiplet of fermions $\widetilde g_a$
which also are color octet.  These particles have no analogue in the
standard model.

 In fact, the restriction \AAcomm\  makes it impossible to explain any
fermion-boson correspondence seen so far in Nature.  A left-handed
quark $q_L$ has as its supersymmetry partner a boson $\widetilde q_L$
with color 3, $I = \half$, and hypercharge $Y = {1\over 6}$.  The
right-handed leptons $\ell_R$ or $\bar \ell_L$ lead to scalar particles
with $I = 0$ and $Y=1$.  The $SU(2)$ doublets of left-handed charged
leptons and neutrinos, such as $E_L = (\nu_e, e_L)$, lead to scalar
particles with $I= \half$, $Y = -\half$.  These quantum numbers are,
curiously, those of the Higgs boson $\phi$.
  But it is difficult to understand
how lepton number could be conserved while $\phi$ obtains a vacuum
expectation value in a scheme where $\phi$ is the partner of $E_L$.

\REF\Nilles{H. P. Nilles, \sl Phys. Repts. \bf 110, \rm 1 (1984).}
\REF\lang{P. Langacker and M.-X. Luo, \sl Phys. Rev. \bf D44, \rm
           817 (1991); P. Langacker and N. Polonsky, \sl Phys.
                     Rev. \bf D47, \rm 4028 (1993).}

   Thus, if supersymmetry is the origin of spin $\half$, one cannot
extend this idea to explain the detailed content of the standard model.
One must, in fact, postulate a new, undiscovered particle as the
partner of each known particle of the standard model.  However, there
are good reasons to believe that Nature is, nevertheless, supersymmetric
at its most fundamental level.  If one believes in the grand unification
of the gauge interactions of the standard model at some very high
momentum scale $m_G$,
 supersymmetry is the
only known way to stabilize the relation $m_\phi \ll m_G$, and is the
most natural explanation of the values of coupling constants observed
at the $Z^0$.   These motivations for supersymmetry are reviewed in
more detail in refs. \Nilles\ and
\lang.   In the remainder of my discussion,
I will use supersymmetry, quite independently of its phenomenological
justification, as a powerful geometrical symmetry which organizes
our conception of space-time.

  I should remark that, just as the supersymmetry
  relation between spin 1 and spin $\half$ particles generates the
spin $\half$ interactions, so the relation between spin $\half$ and
spin 0 generates a set of vertices for the spin 0 fields.  These are
shown in Fig. \Intss(b).   The spin 0 partner of the lepton doublet
$E_L$, for example, acquires a coupling both to the $W$ boson and to
its partner the $\widetilde w$, and also a 4-scalar self-coupling.

\section{Spin $\thalf$ and Higher Supersymmetries}

   From the viewpoint of the theory of spin, one important feature of
supersymmetry is that it provides the missing ingredient in the
discussion of spin $\thalf$ particles given at the end of Section
5.  We argued there that a consistent theory of spin $\thalf$
requires a conserved spin $\thalf$ current $s_{\mu\a}$.  Such a
current would be associated with a global charge
$$        Q_\a = \int d^3x \, s_{0\a} \ .
\eqn\spintcharge$$
We have now seen that such a global charge is consistent only if it is
a supersymmetry charge; then a spin $\thalf$ field can be included in a
field theory only if it couples to the current of supersymmetry.

In such a structure, the spin $\thalf$ field  $\psi_{\mu\a}$ becomes
the gauge field of supersymmetry, and the whole theory acquires a
gauge symmetry
$$        \delta \psi_{\mu\a} = D_\mu \e_\a \ ,
\eqn\psitranse$$
where $\e_\a(x)$ is an anticommuting parameter which defines a local
supersymmetry transformation.  The composition of supersymmetry
transformations gives a translation, and so the field $\psi_{\mu\a}$
should naturally be associated with the gauge field of local
translations, the gravitation field.  Indeed, the massless
spin $\thalf$ particle created by $\psi_{\mu\a}$ is naturally
paired with the graviton in a multiplet with the form of \arels.
The field $\psi_{\mu\a}$ is then called the {\it gravitino}, and the
resulting generalization of general relativity is called
{\it supergravity}.
The Lagrangian which includes gravitation and the natural coupling
of gravity to $\psi_{\mu\a}$,
$$ \L = \sqrt{-g} R  + \half \bar\psi_\mu i\e^{\mu\nu\lambda\sigma}
        \gamma_\nu D_\lambda \psi_\sigma \ ,
\eqn\supergrav$$
can be shown to be invariant under local supersymmetry transformations
which link the fields $g_{\mn}$ and
 $\psi_{\mu\a}$.\Ref\supergrav{D. Z. Freedman, S. Ferrara, and
 P. van Nieuwenhuizen, \sl Phys. Rev. \bf D13, \rm 3214 (1976).}

  In this line of argument, it seems that there could be at most one
spin $\thalf$ field, just as there can be at most one graviton.
However, more general supersymmmetry algebras than \susycomm \ are
possible, and by incorporating them, we may build larger theories.
But this extension also brings in new  restrictions, since the
higher supersymmetry algebras imply still stronger relations among
the couplings of the model.

  The general restrictions of the Coleman-Mandula theorem on the
presence of spin $\half$ changes were worked out by
  Haag, Lopuszanski, and
  Sohnius.\Ref\HLS{R. Haag, J. Lopuszanski, and M. Sohnius,
               \sl Nucl. Phys. \bf B88, \rm 257 (1975).}
These authors showed that, although the restriction we found on the
right hand side of \susycomm\ is absolute, more general theories can be
built by incorporating several supersymmetry charges, each of which
has a square which is the total energy-momentum.  More explicitly,
they allow a set of commutation relations
$$   \big\{ Q^i_\a, Q^{j\dagger}_{\dot \a} \big\} =
 2 \delta^{ij} \sigma_{a\dot\a}^\mu
                       P_\mu \ .
\eqn\susycommN$$
This structure is known as $N$-{\it extended} supersymmetry.
A theory with $N$  supersymmetry charges can be gauged with
one graviton and $N$ gravitinos.  The each of the various supersymmetry
charges pair one gravitino with the graviton and the others with
spin 1 bosons which must also be present in the theory.  These
spin 1 bosons provide a gauge group which does not commute with the
supersymmetry charges $Q^i_\a$, providing the generalization of
\AAcomm.  We can use \susycommN\ to determine the exact particle
content of theories with this higher symmetry.

  To analyze the implications of \susycommN, consider once again the
action of the supersymmetry generators on massless one-particle states
moving in the $\hat 3$ direction.  As in the paragraph below
\susycommml, we can convert the supersymmetry commutation relation
$$   \big\{ Q^i_-, Q^{j\dagger}_{\dot -}\big\} = 2\delta^{ij}
    ( H + P^3)
\eqn\nextsusyl$$
to a set of relations for fermion create and annihilation operators.
Define
$$  {1\over \sqrt{4P^3}} Q^i_- = a^i \ , \qquad
  {1\over \sqrt{4P^3}} Q^{j\dagger}_{\dot -} = a^{j\dagger} \ .
\eqn\newasi$$
The operators $a^i$ are helicity raising operators.
Then the commutation relations \nextsusyl\ become
$$ \big\{ a^i,a^{j\dagger}\big\} = \delta^{ij}  \ .
\eqn\susya$$
These operators build up a multiplet of $2^N$ states connected by
extended supersymmetry.

 The simplest examples of these multiplets occur in theories of
$N=2$ extended supersymmetry.  The simplest representation, which is
built on a state $\ket{p,0}$ of helicity zero, is
$$ \pmatrix{ &  a^1 \ket{p,0} & \cr
          \ket{p,0} & & a^1a^2\ket{p,0}\cr
              & a^2 \ket{p,0} & \cr}
    =
 \pmatrix{ &   \ket{p,\half}& \cr
          \ket{p,0} & &\ket{p,1}\cr
              &  \ket{p,\half} & \cr}  \ .
 \eqn\ntworep$$
This multiplet contains a gauge boson, two chiral fermions, and a
scalar: $(\phi^A, \lambda^{1A}_\a,$ $\lambda^{2A}_\a, A^A_\mu)$.

The $N=2$ multiplet with maximum helicity 2 is
$$\pmatrix{ &   \ket{p,\thalf}& \cr
          \ket{p,1} & &\ket{p,2}\cr
              &  \ket{p,\thalf} & \cr}  \ ,
 \eqn\ntworepg$$
which contains a vector boson, two gravitinos, and the gravitons.
The vector boson may be thought to generate a gauge symmetry unified
with gravity.  However, in this case, the symmetry is only $U(1)$.
To construct higher symmetries within the supersymmetry multiplets,
we must go to higher $N$.

  For $N=4$, one finds a multiplet
$$ \ket{p,-1} \leftrightarrow 4\times \ket{p,-\half} \leftrightarrow
  6\times \ket{p,0}\leftrightarrow 4\times \ket{p,\half}\leftrightarrow
               \ket{p,1}   \ .
\eqn\nfourmult$$
This multiplet is $CPT$ self-conjugate, it contains one vector boson,
4 chiral fer\-mi\-ons, and 6 real scalar bosons.  This is the largest
multiplet for which all fields have spin less than or equal to 1.
The field theory of this multiplet turns out to be quite magical; for
example, its renormalization group $\beta$ function vanishes to all
orders in perturbation theory.\Ref\mandel{S. Mandelstam,
 \sl Nucl. Phys. \bf B213, \rm 149 (1983).}

  One might similarly ask for which values of
  $N$ one finds a multiplet with
a single graviton and no spin higher than 2.   The largest such
multiplet occurs for $N=8$:
$$\eqalign{
 \ket{p,-2} \leftrightarrow 8\times \ket{p,-\thalf} \leftrightarrow
&  28\times \ket{p,-1}\leftrightarrow
  56\times \ket{p,-\half}\leftrightarrow
  70\times \ket{p,0} \cr&\leftrightarrow
  56\times \ket{p,\half}\leftrightarrow
  28\times \ket{p,1}\leftrightarrow
  8\times \ket{p,\thalf} \leftrightarrow
               \ket{p,2}   \ .     \cr}
\eqn\neightmult$$
The multiplet contains 28 gauge bosons. These form the antisymmetric
tensor representation  of $SO(8)$,
 which is also the adjoint representation,  Thus, this theory
naturally contains a unified $SO(8)$ gauge theory.
Unfortunately, this group is not large enough to
contain the standard model
gauge group $SU(3)\times SU(2)\times
U(1)$.  Worse, this $SO(8)$ has vector-like couplings rather than the
chiral couplings which are essential to build weak interaction theory.
Cremmer and Julia\Ref\CJ{E. Cremmer and B. Julia, \sl Nucl. Phys.
       \bf B159, \rm 141 (1979).}
 have worked out the detailed structure of the
$N=8$ supergravity theory and have identified a large  global symmetry
group---a noncompact $E_8$---but so far no one has succeeded in
building a relation between this group and the gauge group of the
standard model.

 Thus, the idea that spin $\half$ arises as the result of a symmetry of
Nature turns out to be a very powerful one, but one which stops
 short of providing a complete theory of the fundamental
interactions.  To build more successful models, we need to add to
supersymmetry some structure of a quite different kind.

\chapter{A Fruitful Blend}

 One of the most remarkable theoretical developments of the past
ten years has been the realization that it is possible to merge the
ideas of the previous two sections  in a fruitful way.   In Section 6,
we studied the idea of building a particle theory by putting a
one-dimensional quantum field theory on a world line.  It is not hard
to imagine a generalization in which one imagines a particle as
a line or ring which sweeps out a two-dimensional  surface in space-time.
In this context, we could build a particle
model by putting a two-dimensional quantum field theory on this surface.
It is not so obvious why this would give an improvement over the
picture we found for world-line theories, or, on the other hand, why
we should stop at two-dimensions rather than studying three- or
four-dimensional objects embedded in space-time.  I can only say that the
two-dimensional case offers just the right balance between freedom and
constraints to allow one to create
 theories with an intriguing amount of structure.
This particular case is known as {\it string theory}, or, with the
inclusion of supersymmetry, {\it superstring theory}.

\section{The Bosonic String}

\FIG\StringG{The dynamics of a periodically connected two-dimensional
            surface is viewed as a particle moving and interacting
          in space-time.}
  The simplest sort of string theory is one in which a particle is a
ring moving through space-time.  The physical picture of particle
motion is that shown on the right-hand side of Fig. \StringG:  The
particle sweeps out a world-surface in the form of a tube.  This
surface may split or branch, and these branches represent particle
interactions.  All of the properties of the particles described
by this motion follow from a quantum field theory on two-dimensional
surface shown on the left.

  In the world-line theory of Section 6.1, we began our discussion by
considering the world-line to be embedded in space-time.  If we
take a similar point of view here, we would describe the string
dynamics by field $X^\mu(s,t)$.
                The arguments of the field
 $(s,t)$ are coordinates on this surface. With a convenient
choice of gauge, the $X^\mu(s,t)$ are free fields whose Lagrangian
is
$$  \int \L  = {T\over 8\pi} \int dsdt  \, \bigl[ \d_\lambda
 X^\mu \d^\lambda
           X_\mu  \bigr] \ ,
\eqn\stringbl$$
with $\lambda = 0,1$. The parameter $T$ has the dimensions of
(mass)${}^2$ and provides a natural length scale for the theory.

Now we can switch our perspective and regard space-time
$X^\mu$ as a set of fields which
lives on the string. Choose the coordinate $s$ to run from 0 to
$2\pi$ around the ring.  Then $X^\mu(s)$ has the Fourier
decomposition
$$  X^\mu(s) = X_0^\mu  + \sum_{n\neq 0} e^{ins} X_n^\mu \ .
\eqn\XsFT$$
Then $X_0^\mu$ is the center of mass position of the string, conjugate
to the total \hbox{4-momentum}
$P^\mu$.  The $X_n^\mu$ are the coordinates of
string oscillations.  These can be quantized as harmonic oscillators
corresponding to running waves moving to the left and to the right
around the ring.  Let $a^\mu_n$ and $\bar a^\mu_n$, $n > 0$, be the
annihilation   operators  corresponding to these two sets of
harmonic oscillators.  Then  the dynamics of the string is neatly
captured in the formula for the energies of the   various string
states.  This is   the relativistic mass formula $P^2 = m^2$, with
$$  m^2 = 2\pi T \Big\{\sum_n  n\bigl( a^{\dagger\mu}_n a^\mu_n +
              \bar a^{\dagger\mu}_n \bar a^\mu_n \bigr) + 2{\cal Z}
                      \Big\} \ .
 \eqn\stringmass$$
The offset $2{\cal Z}$ is the (renormalized) zero-point energy of
the two sets of oscillators.
The relation \stringmass\ between the oscillator excitations
the masses of string states clarifies that the system is at the same
time relativistic and harmonic.

 The mass formula \stringmass\ is reminiscent of the mass formula
in \manystates\ for the particle theory with spin 1.  However, the
two-dimensional field theory substructure provides three further
restrictions.  The first restriction is relatively simple:
The level of excitation in the left-moving
oscillators must be equal to the level of excitation in the right-moving
oscillators.  Otherwise, the coordinate system would rotate around the
ring.   The second restriction is a bizarre one: The number of dimensions
of the space-time in which the string is embedded must be 26.  I will
explain below how this restriction may be relaxed.  The third restriction
precisely fixes the  zero point energy ${\cal Z}$ to the value $(-1)$.

  Ignoring, for the moment, the strange second requirement, we can work
out the lowest mass states in the spectrum of the string.  The ground
state is
$$          \ket{0} \ , \qquad  m^2 = -4\pi T \ .
\eqn\tachy$$
This state is an unphysical scalar tachyon, and this also must be
eliminated in an improved theory.  The first excited states are
$$    a^{\dagger\mu}_1\ket{0}\ , \quad
    \bar a^{\dagger\mu}_1\ket{0}\ , \qquad    m^2 = -2\pi T \ .
\eqn\firstex$$
However, these states  are eliminated by the requirement that the
left- and right-moving oscillators have the same degree of
excitation.  Thus, the next physical state of the theory is
$$    a^{\dagger\mu}_1 \bar a^{\dagger\n}\ket{0}\ ,
    \qquad    m^2 = 0\ .
\eqn\secondex$$
This multiplet contains an exactly massless spin 2 particle---the
graviton.

 From the discussion of  Section 5, we ought to be suspicious
  \FIG\FillHole{Relation between the amplitude for emission of a
        string state and an operator expectation value in the
           world-surface.}
that this spin 2 particle is defined consistently, so that its
unphysical components are not produced in scattering processes.
However, in the string theory, one can prove that the unphysical
production amplitudes naturally cancel.  The strategy of this proof
is geometrical, and is illustrated in Fig. \FillHole.   The emission
of any string state occurs through a world-surface of the form
shown on the left of the figure,  with a long pipe branching off of
the main surface.  If we deform the pipe to be very long and thin,
we can  replace the pipe with a pointlike perturbation of the
world surface, which can be represented by a local operator.
Each particular particle state of the string corresponds to a
different boundary condition at the end of the pipe, and therefore,
through this construction, to a different local operator.  For the
particle state \secondex, the corresponding local operator is the
energy-momentum tensor $T^{\mu\nu}(s,t)$ on the world surface.
This is a conserved tensor, and so the unphysical components of the
graviton cancel out naturally.

  Actually, this argument is reveals only a part of the deeper
substructure of string theory.  In higher dimensions, as we have
seen, the energy-momentum tensor is the highest rank conserved
tensor possible.  However, in two dimensions it is possible to
have extremely large geometrical symmetry groups and correspondingly
high-rank conserved tensors. The special symmetry involved is familiar
from the theory of classical partial differential equations: In
two dimensions, equations which do not have an intrinsic scale, such as
the Laplace equation, are solvable by conformal mapping.  Under
certain conditions, this
symmetry of conformal mapping survives into the quantum theory.
For free fields, these conditions precisely restrict the total number
of fields in the theory and the zero-point energy, in just the manner
described below \stringmass.   Since a surface branching into a long
thin tube is related to a surface with a small hole by a conformal
transformation, the relation shown in Fig. \FillHole \ requires no
approximation and applies to every possible string state.  This
means that the interactions of a string are uniquely determined
by its spectrum, a profound generalization of the constraints of
gauge invariance.  Similarly, the cancellation of
unphysical states that we found for the graviton generalizes to
the full string spectrum:
  By using the relation between the higher states of the string
spectrum and the higher conserved tensors of the
two-dimensional theory, one can show that all
 negative norm excitations created by the
 operators $a^{\dagger 0}_n$  cancel out of
scattering matrix elements.\Ref\NoGhost{J. Scherk, \sl Rev. Mod. Phys.
            \bf 47, \rm 123 (175).}

\section{Decorated Strings}

  The theory in which only the space-time coordinates $X^\mu(s,t)$
live on the world surface has some wonderful mathematical properties
but also contains awkward features---a tachyonic particle and the
restriction to 26 dimensions.  To ameliorate these problems, we might
try to put a different two-dimensional field theory onto the string
world surface.  Following the approach of Section 5.2, we can add
anticommuting coordinate $\psi^\mu(s,t)$.  In order for the negative
norm excitations created by $\psi^0$ to be cancelled, we require a
higher symmetry which incorporates both conformal invariance and
supersymmetry.  Thus, to add spin to the string in the manner of
Section 5.2, we must already add two-dimensional supersymmetry.
What do we get back in return?

   Comparing this construction to that in Section 5.2, we see one
new feature:  The field $\psi^\mu$ is a function of the coordinate
$s$ which runs around the ring, and we must fix the boundary
condition to be imposed on these fields.  The simplest choice is
periodic boundary conditions.  With this choice, the zero point
energy turns out to be $\Z = 0$ and so the $n=0$ Fourier
components of the $\psi^\mu$ link a multiplet of massless particles.
In the quantum theory of the string, the operators corresponding
to these modes satisfy
$$          \big\{\psi^\mu_0,\psi^\nu_0 \big\} = - \half g^\mn \ ,
\eqn\theanticommtwo$$
just as we found for the one-dimensional case in \theanticomm.
The physical interpretation is the same:  These strings are
Dirac fermions in space-time.   In the discussion below, I will
denote states of this Dirac fermion multiplet as  $\ket{\alpha}$.

   In the same string theory, it is consistent to have other
strings for which the boundary condition on $\psi^\mu(s)$ is
different.  One may consider, for example, antiperiodic  boundary
conditions:  $\psi^\mu(2\pi) = -\psi^\mu(0)$.    Now the Fourier
expansion of $\psi^\mu(s)$ has the form
$$ \psi^\mu(s) =  \sum_{n \geq 0} e^{i(n+\half) s} \psi_{n+\half}^\mu
                     + c.c.
\eqn\psisFT$$
The zero point energy for these strings is $\Z = -\half$, so the
states
$$            \psi^{\dagger\mu}_\half \bar \psi^{\dagger\mu}_\half
                        \ket{0}
\eqn\thefgrav$$
form a multiplet of massless particles which include the graviton.
In fact, $\psi^\mu_\half$ and $\psi^{\dagger\mu}_\half$ have the
same operator relations as the operators $\theta^\mu$ and
$\bar\theta^\mu$ in \thenewHam.  Remarkably, string theory allows
particles acted on by these operator to coexists with particles
acted on by \theanticommtwo.   The price of this coexistence is
equally remarkable; it is that states with an even number of fermions
in either the left- or right-moving sector cancel out of the $S$-matrix.
This removes the tachyon $\ket{0}$, and also converts the multiplet
$\ket{\alpha}$
acted on by \theanticommtwo\ from a Dirac to a chiral fermion.

   For either or both choices of boundary condition, the constraint on
the total number of fields $X^\mu$ and $\psi^\mu$ is that $\mu$ should
run over 10 dimensions.  However, it is possible to lower this number
to 4 dimensions, or any other convenient value,
 by decorating the string world surface with more free fields, or with
interacting fields which satisfy the constraints needed for conformal
invariance.  The zero-point energy depends on the detailed collection
of fields, and may be different for the left- and right-moving
sectors.  For example, one can build a theory with additional
right-moving antiperiodic fermions $\bar \chi^i$, $i= 1,\ldots, n$,
in such a way that
the total zero point energy is $\Z = -\half$ from the left-moving
sector and $\bar\Z = -1$ from the right-moving sector.  Then the
state
$$            \psi^{\dagger\mu}_\half \bar\chi^{\dagger i}_\half
                   \bar\chi^{\dagger j}_\half \ket{0}
\eqn\avect$$
is an allowed particle of the theory.  This state is massless, has
spin 1, and transforms as an antisymmetric tensor of $SO(n)$.  It is,
in fact, an $SO(n)$ gauge boson.

   If the content of the string theory is properly arranged,
states which acquire a vector character from a left-moving
excitation created by $\psi^{\dagger\mu}_\half$ can be naturally
paired with states for which the left-moving sector is a spinor
state $\ket{\alpha}$.  For example, the same theory which contains
the state \thefgrav\ will also contain
$$             \bar \psi^{\dagger\mu}_\half
                        \ket{\alpha} \ .
\eqn\thefgrav$$
This particle carries spin $\thalf$ and is, in fact, the gravitino
partner of the graviton given above.  Similarly, the same theory which
contains the state \avect\ will also contain
$$        \bar\chi^{\dagger i}_\half
                   \bar\chi^{\dagger j}_\half \ket{\alpha} \ ,
\eqn\avect$$
the spin $\half$ supersymmetry partner of the gauge boson.  In this way,
it is straightforward to construct string theories which have a
supersymmetric spectrum, and, by extension, a full set of supersymmetric
interactions.

 \REF\STMod{M. Dine, ed. \sl String Theory in Four Dimensions. \rm
          (North-Holland, Amsterdam, 1988).}
\REF\STModtwoa{B. Schellekens, ed. \sl Superstring Construction. \rm
          (North-Holland, Amsterdam, 1989).}
\REF\STModtwo{L. J. Dixon in \sl From Actions to Answers, \rm T. De
            Grand and D. Toussaint, eds. (World Scientific, Singapore,
                            1990).}

   From this point, the possibilities are limited only by one's
imagination in assembling two-dimensional field theories with which to
decorate the string world surface.  A large number of model-building
strategies for string theory are described in refs. \STMod--\STModtwo.
 Using these strategies, one can build a wide variety of theories,
some of which might even resemble the standard model.  All of these
theories, or at least all interesting ones, require spin $\half$ or
some generalization as an input to build the two-dimensional theory,
but then recover spin $\half$, and often supersymmetry as well, in
its spectrum of particles in space-time.

\chapter{Conclusions}

Though we have found no definite answer to the question posed in the
introduction, we have found ourselves led through a wonderful tangle
of speculations on the deep structure of Nature.  Is spin constructed
or is it fundamental?  Is it the requirement of symmetry?  In the
furthest flights we have taken, it seems that space-time itself is
too restrictive a notion, and that we must generalize this in order
to gain a full appreciation of spin.  In any case, there is no doubt
that spin must play a central role in unlocking the  mysteries of
fundamental physics.

\ack

I am grateful to David Burke and Lance Dixon, for suggesting this
series of lectures, and to Daniel Schroeder, for many discussions
of the basics of quantum field theory.

\endpage
\refout
\endpage
\figout
\endpage
\end